\documentclass[11pt]{iopart}
\pdfoutput=1
\usepackage{fullpage}

\usepackage{iopams,graphicx,color}
\usepackage{setstack}
\usepackage[pdftex,breaklinks=true,bookmarks=false,colorlinks=true,citecolor=blue,urlcolor=blue,linkcolor=black]{hyperref}

\newcommand{\psm}{p_{\sigma,\mu}}

\begin{document}
 
\title[Entanglement in RVB and quantum dimer states]{Entanglement in gapless resonating valence bond states}
 
\author{Jean-Marie St\'ephan$^1$, Hyejin Ju$^2$, Paul Fendley$^1$ and Roger G. Melko$^{3,4}$}

\address{$^1$ Physics Department, University of Virginia, Charlottesville, VA 22904-4714}

\address{$^2$ Department of Physics, University of California, Santa Barbara, CA, 93106-9530}

\address{$^3$ Department of Physics and Astronomy, University of Waterloo, Ontario, N2L 3G1, Canada}

\address{$^4$ Perimeter Institute for Theoretical Physics, Waterloo, Ontario N2L 2Y5, Canada}

\eads{\mailto{jean-marie.stephan@virginia.edu}, \mailto{ju@physics.ucsb.edu}, \mailto{fendley@virginia.edu}, \mailto{rgmelko@uwaterloo.ca}}

\date{\today}
\begin{abstract}

We study resonating-valence-bond (RVB) states on the square lattice of spins and of dimers,  as well as $SU(N)$-invariant states that interpolate between the two. These states are ground states of gapless models, although the $SU(2)$-invariant spin RVB state is also believed to be a gapped liquid in its spinful sector. We show that the gapless behavior in spin and dimer RVB states is qualitatively similar by studying the R\'enyi entropy for splitting a torus into two cylinders.  We compute this exactly for dimers, showing it behaves similarly to the familiar one-dimensional log term, although not identically. We extend the exact computation to an effective theory believed to interpolate among these states. By numerical calculations for the $SU(2)$ RVB state and its $SU(N)$-invariant generalizations, we provide further support for this belief. We also show how the entanglement entropy behaves qualitatively differently for different values of the 
R\'enyi index $n$, with large values of $n$ proving a more sensitive probe here, by virtue of exhibiting a striking even/odd effect.

\end{abstract}
\maketitle

\tableofcontents
\vfill\eject

\hypersetup{linkcolor=red}

\section[Introduction]{Introduction}

\label{sec:introduction}

A vast amount of recent theoretical and experimental effort has been devoted to the study of and search for spin liquids. A spin-liquid phase typically exhibits exotic behavior such as quasiparticles 
with fractionalized quantum numbers, or spin-charge separation. Such behavior is manifestly non-perturbative, and so requires advanced theoretical tools to analyze. 

A vast amount of recent theoretical effort has been devoted to understanding how the entanglement properties of a particular state illuminate its physical properties. Since a major lesson of condensed matter physics in recent decades was that a phase often can be characterized by a model state, a feature of studying entanglement is that states can be studied directly, without needing to know a specific Hamiltonian. 
Since studying entanglement properties of a model state is often both analytically and numerically more tractable than analyzing a model Hamiltonian, much recent progress has been made in condensed matter physics from the former.

The behavior of quantities such as the R\'enyi entropy thus can be used to explore spin-liquid states directly. The quantum spin liquids best characterized theoretically are { gapped}, where a gauge symmetry typically emerges.
An important and useful concept applicable here is topological order, which relates the emergent gauge symmetry to a topological degeneracy, and the related concept of { topological entropy} \cite{HIZ,KP,LW}. 
However, many experimental systems currently identified as good candidates for spin liquids appear to be {\it gapless}.  Compared to their gapped counterparts, these phases are less amenable to simple unifying attributes such as topological order.  Although it is likely that other quantities, related perhaps to correlations or manifestations of long-range entanglement, can be used to characterize gapless phases, few viable examples of {\it models} -- particularly those tractable by large-scale numerical simulation -- are known to contain gapless spin liquids.

We study a particular family of states, resonating-valence-bond (RVB) states on the square lattice, that are gapless and exhibit spin-liquid-type behavior \cite{RokhsarKivelson,RVB1,RVB2}. Our main (but not only) tool is to analyze a subleading term in the R\'enyi entropy particularly useful for characterizing gapless systems \cite{Ju2012}. The R\'enyi entanglement entropies $S_n$ are defined as
\begin{equation}
S_n=\frac{1}{1-n} \ln {\rm Tr}\, \rho_A^n\ , \label{RenyiEq}
\end{equation}
where $\rho_A$ is the reduced density matrix for a region $A$, and the von Neumann entropy $S_1$ is defined as the limit $n\to 1$.
The term we analyze describes the entanglement between two regions formed by cutting a torus into two cylinders. This geometry is useful for studying gapless systems
because the length of the boundaries separating the cylinders is independent of the area of the cylinders. Because of the long-range correlations, the entanglement
entropy of gapless systems will be sensitive to varying the size of the cylinders. 
This dependence was studied for three completely different gapless systems in \cite{Ju2012}, and argued to have universal behavior. One of the results of this paper is to provide additional evidence in support of this assertion.

One model with a gapless RVB ground state, the square-lattice quantum dimer model, is quite amenable to an analytic treatment. For example, all equal-time correlators diagonal in the dimer basis are those of a classical dimer model \cite{RokhsarKivelson}. These can be computed exactly directly in the dimer model by using Pfaffian techniques \cite{Kasteleyn,Fisher,FisherStephenson}, or in the continuum limit by utilizing a two-dimensional classical free-boson field theory \cite{Fradkinbook,Henley}. Such exact computations have been done for entanglement quantities as well \cite{FradkinMoore,Hsu2009,Shannonee,Oshikawa,Zaletel,Stephan2011}. In this paper we extend these methods to compute the two-cylinder entanglement entropy for the quantum dimer model exactly in the scaling limit. For R\'enyi index $n$ greater than a specific (non-universal) value $n_c$, it is simply given by a ratio of cylindrical partition functions. From this we find for example that the striking even-odd effect 
observed numerically in \cite{Ju2012} for 
the $SU(2)$-invariant RVB state was not a finite-size artifact; it also occurs for the quantum dimer model, and our computation illuminates its origin.

One lesson from our results is that universal entanglement properties can be qualitatively different for values of the R\'enyi index $n$. For $n>n_c$, the two-cylinder entanglement entropy depends strongly on whether the cylinders are of odd or even length. This is not shocking from the point of view of the quantum dimer state, because the correlators exhibit similar behavior. However, at the von Neumann point ($n\to 1$), no such even-odd effect is possible. Indeed, the shape of the entanglement entropy has to be a concave-down function of $y$, as required by the strong subadditivity property \cite{Strongsubadditivity}. Other values of $n$ are not constrained by subadditivity, and we will derive this even/odd effect explicitly for the dimer RVB state. 
Thus only for $n>n_c$ is the R\'enyi entropy sensitive enough to include this effect. 
This effect, whereby a R\'enyi entropy may exhibit qualitatively different behavior for some higher $n$, is reminiscent of the manifestation of finite-temperature criticality in R\'enyi entropies for $n>1$, but not $n=1$ \cite{RenyiXing}. 

It is possible to interpolate between the $SU(2)$-invariant RVB state and the quantum dimer state by considering $SU(N)$-invariant generalizations of the former; many aspects of the quantum dimer state then can be extracted from the $N\to\infty$ limit. Such a correspondence can be pushed further: Ref.~\cite{Damle} defined a cluster expansion of the RVB ``loop gas''\cite{Sutherland_loops} in terms of interacting dimers in the large $N$ limit. One recovers to first non-trivial order the interacting generalization of the dimer model studied in \cite{Alet_dimers1,Alet_dimers2}. Most strikingly, the critical exponent $\alpha$ extrapolated to $N=2$ turned out to be in good agreement with the previous numerical studies \cite{RVB1,RVB2}. This expansion also provides some insight in the study of analogous RVB states in three-dimension \cite{AAM,AMPMJ}.

Whereas one cannot do exact computations directly in these models outside of the dimer case, the field theory computation can be extended to a full line of critical points by varying the stiffness of the classical boson. This line of critical points likely describes the scaling limit of the interacting dimer model \cite{Alet_dimers1,Alet_dimers2,Damle}.  We expect that the $SU(N)$ version studied here also will have the same universal behavior, and by comparing our quantum Monte Carlo results to the expansion developed in \cite{Damle}, we provide strong evidence that the scaling limit of the $N=2,3,4,5$ theories indeed are points on this line. Moreover, we provide consistency checks that the two-cylinder entanglement entropy also behaves universally for $N=2,3,4$. Our results therefore strongly imply that at least the gapless part of the square-lattice RVB state behaves qualitatively similar to the quantum dimer model; there is no phase transition as $N$ is varied.

The paper is organized as follows:
In section\ \ref{sec:RVB}, we review the definition and important properties of the RVB states. We derive our general result for the R\'enyi entanglement entropy of many critical Rokhsar-Kivelson type states, including the quantum dimer state, in section \ref{sec:shape_general}. Namely, for $n>n_c$ the R\'enyi entropy is given by a simple ratio of 2d classical partition functions.   Section \ref{sec:dimers_entanglement} explores the consequences of this result in the simplest case of the square lattice quantum dimer state. We also discuss the emergence of a strong even-odd effect, similar to that first observed in \cite{Ju2012}. We then turn our attention in section \ref{sec:correlations} to the $SU(N)$ RVB state.  By quantum Monte Carlo simulations, we show that while the spin-spin correlations decay exponentially, the dimer-dimer correlations are critical. The exponent approaches the dimer exponent ($\alpha_D=2$) as $N$ increases, and agrees well with a previous large $N$ computation \cite{Damle} and 
earlier numerical results at 
$N=2$ \cite{RVB1,RVB2}. We also present further evidence of the underlying Coulomb-gas structure for all $N$.
Finally, in section \ref{sec:rvb_entanglement} we present numerical simulations of entanglement for the $SU(N)$ RVB wave function. The behavior qualitatively resembles that of the dimer RVB state, and therefore provides strong evidence of universal behavior, in agreement with our main result of sec.~\ref{sec:shape_general}.

\section{The RVB states and their equal-time correlators}
\label{sec:RVB}

A paradigm for a spin liquid is the two-dimensional resonating-valence-bond (RVB) state \cite{Anderson}. Much effort has gone into its study, since the classic work in the late '80s \cite{LDA,Sutherland_loops}. 
In this section we introduce and review some of the known properties of these states.

\subsection{The $SU(2)$ RVB state}
 A valence bond is a spin singlet, and a valence-bond state is one where each spin forms a valence bond (i.e.\ a singlet) with one other spin. The nearest-neighbor $SU(2)$ RVB wavefunction is the equal-amplitude superposition of all nearest-neighbor valence-bond (or singlet) states of spin-1/2 particles fixed at the sites of some lattice. For the ${\cal N}$-site square lattice, the $SU(2)$ RVB state is
\begin{equation}
| \Psi \rangle = \sum_{\cal C} |V_{\cal C} \rangle\ ,
\label{RVBdef1}
\end{equation}
where
\begin{equation}
 |V_{\cal C} \rangle = \frac{1}{2^{{\cal N}/4}} \prod_{i=1}^{{\cal N}/2} \big( | \!\uparrow_i \downarrow_{j_{i}} \rangle  - | \!\downarrow_i \uparrow_{j_i} \rangle  \big)
\label{RVBdef2}
\end{equation}
so each spin $i$ on one sublattice is in a singlet with one of its nearest neighbors $j_i$ on the other sublattice. A valence bond between neighbors $i$ and $j$ can be labeled by a dimer on this link between $i$ and $j$, and so each valence-bond state can be viewed as a dimer configuration ${\cal C}$ with exactly one dimer touching each site. Our computations probe a state directly and so do not require a Hamiltonian. However, it is worth noting that this $SU(2)$ RVB state is the exact ground state of a quantum spin Hamiltonian with local interactions \cite{Cano}.

This RVB state breaks no symmetry, and so is a natural candidate for a state exhibiting spin-liquid behavior.  For example, if the two-dimensional square lattice is periodic (i.e.\ is a torus) and has an even number of sites in both directions, there is a winding number for each direction. These are defined simply by counting the dimers around a closed path going straight around a cycle, weighting each dimer by $+1$ for an even number of steps along the path, $-1$ for an odd number of steps, with no contribution from empty links. An RVB state can be defined for each value of the winding numbers, and 
each will be the ground state of a local Hamiltonian. The number of ground states will then depend on the number of non-trivial cycles, a basic characteristic of topological order. 

For the square lattice, extensive numerical analysis indicates that dimer-dimer correlators in the RVB state decay algebraically \cite{RVB1,RVB2}. A theorem of Hastings guarantees that when any local operators have an algebraically decaying correlator, any local Hamiltonian with this as a ground state must be gapless \cite{Hastings_thm}. Thus the square-lattice RVB state could only describe a gapless system. However, it has long been known, both from numerics and strong analytic arguments, that spin-spin correlators are exponentially decaying \cite{LDA}. Coupled with the fact that no local order parameter has been found \cite{RVB1,RVB2}, it seems very possible that there is a spin gap, and that the state effectively behaves as a spin liquid. In this paper we do not address this question directly. Instead we focus on the gapless behavior of nearest-neighbor RVB states on the square lattice.

\subsection{The dimer RVB state}

 One key tool is to study the {\em dimer RVB state} \cite{RokhsarKivelson}. Here one essentially forgets the underlying spins: the only degrees of freedom are the nearest-neighbor dimers. Each configuration of dimers is a linearly independent basis element of the Hilbert space of the quantum dimer model, and these dimer ``states'' span the Hilbert space. As above, a dimer configuration ${\cal C}$ has exactly one dimer touching each site of the lattice, i.e.\  the dimers are close packed and hard-core. The RVB state for dimers is the equal-amplitude sum over all these states: 
\begin{equation}
| \Psi_D\rangle= \sum_{\cal C} |D_{\cal C} \rangle\ .
\end{equation}
Different dimer configurations are defined to be orthogonal:
\begin{equation}\label{eq:ortho}
\langle D_{\cal C} | D_{{\cal C}^\prime} \rangle = \delta_{{\cal C},{\cal C}^\prime}\ .
\end{equation} 
The ``RK'' Hamiltonian \cite{RokhsarKivelson} is a simple local Hamiltonian with this dimer RVB state as its ground state.

One very convenient feature of the dimer RVB state is that any correlator diagonal in the dimer basis can be computed exactly, because they are exactly those of the two-dimensional classical dimer model. This follows immediately from the definitions of the state and the inner product. For example, the normalization of the dimer RVB state is the partition function of the classical dimer model, the number of distinct dimer configurations:
\begin{equation}
\langle \Psi_D| \Psi_D\rangle = \sum_{\cal C} \ 1\ =\ Z_{\rm dimer} \ ,
\end{equation}
which behaves asymptotically on the square lattice as $\sim (1.79162\ldots)^{\cal N}$ \cite{Kasteleyn}.
 A classic result of classical statistical mechanics is that any correlator in the classical dimer model on any planar lattice can be expressed as a Pfaffian \cite{Kasteleyn,Fisher}. This is equivalent to saying that the model can be rewritten in terms of free fermions \cite{Samuel}. For the dimer-dimer correlation function, the asymptotic behavior can easily be obtained using this expression. For the square lattice long ago this was shown to be algebraically decaying:
\begin{equation}
  C_{dd}(\mathbf{r}_1,\mathbf{r}_2) \sim C\left|\mathbf{r}_1-\mathbf{r}_2\right |^{-\alpha} \ ,
 \label{dimerdimer}
 \end{equation}
with exponent $\alpha=2$ \cite{FisherStephenson}. The coefficient $C$ depends on both the direction and the parity of the number of steps along any path connecting the two dimers; in some special cases it vanishes and so subleading terms dominate. In fermionic language, the dimer creation operator is a fermion bilinear, and so is of dimension 1. Algebraic decay occurs only for bipartite lattices; for non-bipartite lattices, correlators are exponentially decaying, a fact crucial in showing that the triangular-lattice quantum dimer model is in a gapped RVB phase, i.e.\ has topological order \cite{Moessner}.

It is also possible to generalize this construction, with ``enforced'' orthogonality (\ref{eq:ortho}), to other models. For example starting from a classical six-vertex model, we get a quantum six-vertex wave function \cite{QuantumLifshitz}. This wave function also has algebraically decaying correlations, with a continuously varying exponent $\alpha$ as a function of the vertex weights on its critical line. The classical model, although in general more complicated than the dimer model, is nevertheless integrable. It is also constructed with a similar set of constraints, the fully packed and hardcore constraints being replaced by the ice rule. All the analytical results we will obtain for the entropy of quantum dimers can be straightforwardly generalized to the quantum six-vertex case.

The inner product is an important difference between the dimer and $SU(2)$ RVB states.
In the RVB state (\ref{RVBdef1},\ref{RVBdef2}), the inner product is the usual one for spins: each different spin state in the $S^z$ diagonal basis is orthogonal to the others. As is easily shown \cite{Sutherland_loops}, this means that different valence-bond states in the spin model are {\em not} orthogonal. There is a convenient loop-gas description to represent their inner product. One places the two dimer graphs on top of each other, obtaining the ``transition'' graph. Because of the requirement that there be exactly one dimer per site, this means that the transition graph is comprised of closed loops. The shortest loops are of length two, when both graphs have a dimer on the same link. Using the spin inner product then gives \cite{AWSVBSQMC},
\begin{equation}\label{eq:nonortho}
\langle V_{\cal C} | V_{{\cal C}^\prime} \rangle = 2^{n_l - {\cal N}/2}\ ,
\label{RVBinner}
\end{equation}
$n_l$ being the number of loops in the transition graph.
Despite this overlap between dimer configurations making analytic computations much more difficult, its presence is convenient in the algorithm used for computing the second R\'enyi entropy using quantum Monte Carlo simulations \cite{swap},
since ``swapped'' basis configurations (described in Section~\ref{sec:su2_numerics}) will always contribute a non-zero expectation value, even with a naive QMC updating algorithm.

Despite this important difference between the dimer and $SU(2)$ RVB states, their obvious similarity makes it plausible that the dimer-dimer correlator will behave similarly in the two cases. Further evidence for this fact is that rewritten in loop language, the spin-spin correlator corresponds to the probability that the two spins are on the same loop. This decays exponentially for loop models with weight per loop greater than two \cite{Nienhuis}, and the inner product (\ref{RVBinner}) amounts to a weight per loop of 4 \cite{LDA}.  This exponential decay arises because the classical partition function is dominated by configurations with many short loops. Thus correlators in the $SU(2)$ RVB state are dominated by short-loop configurations, i.e.\ the dimers. Whereas the presence of longer loops renormalizes the exponent $\alpha$, it seems very likely that dimer-dimer correlators in the RVB state will decay algebraically like they do for the dimer RVB state.
 
This expectation was convincingly demonstrated numerically \cite{RVB1,RVB2}. The dimer-dimer correlator for dimers was found to be of the form (\ref{dimerdimer}) with
exponent close to $\alpha \approx 1.2$. This is an even weaker falloff than in the dimer RVB state, so the presence of the loops in the inner product of the $SU(2)$ RVB state only makes this correlator longer-ranged. By the theorem of \cite{Hastings_thm}, algebraic decay of local operators in a two-dimensional ground state requires that any local Hamiltonian with this ground state be gapless. Thus not only is the quantum dimer model with RK Hamiltonian gapless,  the spin Hamiltonian of Ref.~\cite{Cano}, whose exact ground state is the $SU(2)$ RVB state, must be as well.

\subsection{The $SU(N)$ RVB state}

It is possible to interpolate between the dimer and $SU(2)$  RVB states, either by dressing the lattice itself \cite{RMS}, or by modifying the $SU(2)$ symmetry to $SU(N)$. In the $SU(N)$ RVB state, the ``spins'' have $N=2S+1$ components, and an $SU(N)$ singlet between sites $i$ and $j$ is given by
\begin{equation}
\label{eq:rvb}
  [i,j]=\frac{1}{\sqrt{2S+1}}\sum_{m\in \{-S,-S+1,\ldots,S\}}^{}(-1)^{m-S}\,|m\rangle_i \otimes |-m\rangle_j\ .
 \end{equation}
When $i$ and $j$ are nearest neighbors, such a singlet can be viewed as a dimer occupying the link between $i$ and $j$,  as in the $SU(2)$ case.
The nearest-neighbor RVB wave function is then defined as an equally weighted superposition of these valence-bond states (\ref{RVBdef1}) as before, with 
 \begin{equation}
 |V_{\cal C} \rangle = \frac{1}{2^{{\cal N}/4}} \prod_{i=1}^{{\cal N}/2} 
 [i,j_i]  \ .
 \label{eq:vbbasis}
\end{equation}
For $SU(N)$ valence-bond states, the orthogonality relation becomes
\begin{equation}
\label{eq:overlap}
 \langle V_{\cal C}|V_{{\cal C}^\prime}\rangle= (2S+1)^{n_{l}-{\cal N}/2}\ ,
\end{equation}
where $n_l$ is again the number of loops in the transition graph formed by superimposing the two dimer configurations ${\cal C}$ and ${\cal C}^\prime$. Notice that $n_l$ can be at most ${\cal N}/2$, and this happens only when all the loops are trivially flat, i.e when the two dimer configurations are identical. We have however
\begin{equation}
 \langle V_{\cal C}|V_{{\cal C}^\prime}\rangle=\delta_{{\cal C},{\cal C}^\prime}
 \end{equation}
 in the limit $S\to \infty$ (or equivalently $N \to \infty$). As a result, any correlation calculated in the limit $N\to \infty$ will be identical to that in the corresponding dimer model. It is however important to stress that this argument strictly only holds at the level of correlation functions.

We will present in this paper a variety of results for the $SU(N)$ case, and show that there is no sign of a phase transition as $N$ is varied, thus giving further evidence for the similarity in behavior between the  $SU(2)$ and dimer RVB states. We will see, for example, that for $N=2,3,4,5$ the dimer-dimer correlator is algebraically decaying, with exponent depending on $N$.

Given the changing exponents, the $SU(N)$ RVB states are obviously not identical to that for dimers, even in the scaling limit. In fact, we will present strong evidence that equal-time correlators for the dimer and $SU(N)$ RVB states can all be written in terms of a free boson, or Coulomb gas. 

\subsection{The Coulomb-gas approach to classical dimers}
\label{sec:cg}

A well established and very useful fact about the classical dimer model is that long-distance properties of the dimers are described by a free bosonic field theory, or Coulomb gas. This can be seen either by taking the continuum limit of the fermionic description (see e.g.\ \cite{FMS}) and then bosonizing the fermions, or by rewriting the dimers in terms of a discrete ``height'' degree of freedom \cite{Heights_1,Heights_2,Henley}. We follow the latter approach, because a careful treatment of the boundary conditions is necessary to obtain our results for the two-cylinder R\'enyi entropy. 

\begin{figure}[ht]
 \begin{center}
 \includegraphics[scale=0.8]{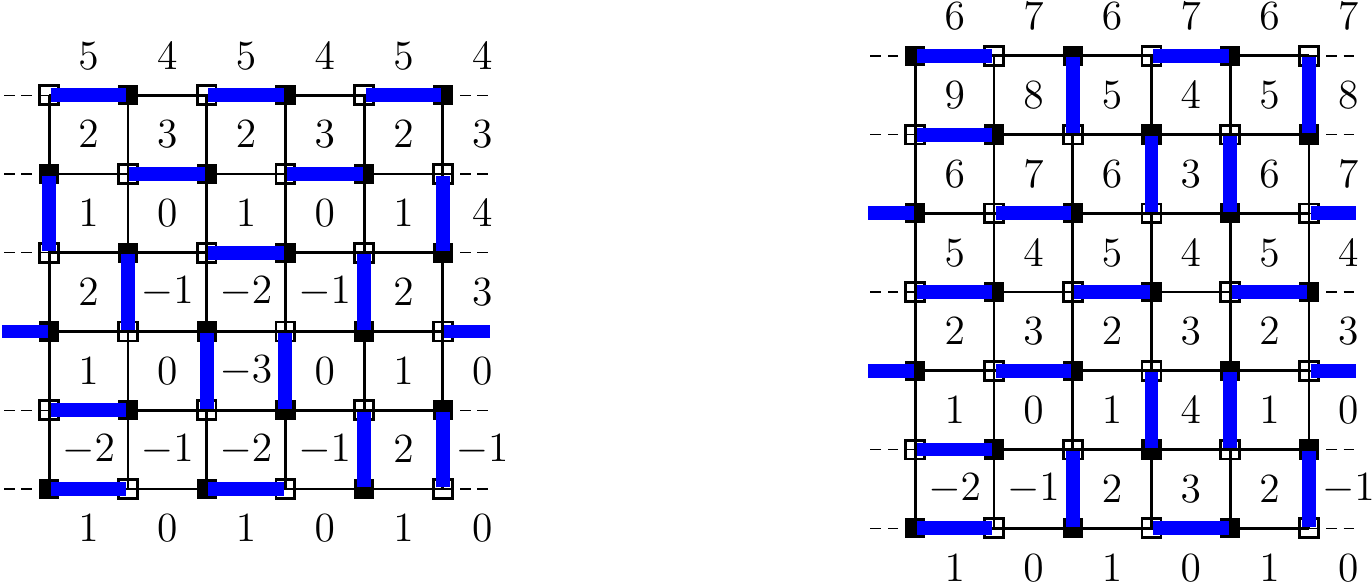}
 \end{center}
\caption{Illustration of the height shift between the upper and lower boundaries in the odd case.  \emph{Left:} Even $L_y$ case. The coarse-grained height difference between top and bottom is $\Delta h=\langle h_{\rm top}\rangle-\langle h_{\rm bottom}\rangle=(\pi/2)(9/2-1/2)=2\pi$.  \emph{Right:} Odd $L_y$ case. We have $\Delta h=(\pi/2)(13/2-1/2)=3\pi$.}
\label{fig:height_shift}
\end{figure}

The first step in the mapping of classical dimers on to the Coulomb gas is to associate an integer with each face of the lattice, as illustrated in fig.~\ref{fig:height_shift}.
These integers are defined by the following construction. Set value on one face (say the lower leftmost one) to be zero. Then, turning counterclockwise around a site of the even (resp.\ odd) sublattice represented by black squares (resp.\ white squares), the integer changes by $-3$ (resp.\ $+3$) if it crosses a dimer, $+1$ (resp.\ $-1$) otherwise. Since by construction these integers are different on adjacent sites, in the bulk one defines a smoother height variable $h(i,j)$ on the sites of the lattice, by averaging the values of the integers on four faces surrounding each point (and then for convenience's sake multiplying by $\pi/2$). Then the states where the dimers are arranged in columns correspond to a constant value of the height.

An important subtlety in this mapping is the boundary conditions, which will play a crucial role in understanding the even/odd effect in the two-cylinder R\'enyi entropy.
With periodic boundary conditions and an even number of sites around this cycle (as in the left and right sides of the figure \ref{fig:height_shift}), note that means that heights can jump by $2\pi$ times an integer on from one side to the other. 
With open boundary conditions, as we have at the ends of the cylinder (the top and bottom of the figure), the natural definition of the boundary height comes from averaging the integers along the boundary. 
It is then easy to check that for a cylinder of even $L_x$ circumference and length $L_y$,  the height difference between top and bottom satisfies
\begin{equation}
 \Delta h=h(x,L_y)-h(x,0)=\left\{
 \begin{array}{ccc}
  2\pi w \quad,\quad w\in \mathbb{Z}&&L_y \;\;{\rm even}\\
  2\pi (w+1/2) \quad,\quad w\in \mathbb{Z}&&L_y \;\;{\rm odd}
 \end{array}
 \right.
\end{equation}
Even/odd effects have long been studied in the dimer model \cite{Ferdinand,Dimers_all,Ruelledimers}, but to our knowledge they have never been interpreted in the context of the height mapping. 
 
The standard Coulomb-gas arguments \cite{Nienhuis} then imply upon coarse-graining the height variable $h(i,j)$ into a continuous field, the scaling limit of the classical dimer model is described by a free scalar field $h(x,y)$ with Euclidean action
 \begin{equation}\label{eq:free_field_bis}
  S_\kappa[h]=\frac{\kappa}{4\pi}\int \left(\nabla h\right)^2 dx dy\ .
 \end{equation}
The coupling $\kappa$ is often known as the ``stiffness''; we show below that for the dimer case it is set to $\kappa_D=1/2$. This free-boson field theory is one of the simplest examples of a two-dimensional conformal field theory \cite{Ginsparg}. It is ubiquitous in two-dimensional statistical models, as well as in condensed matter, where it describes Luttinger liquids.  In order to be consistent with periodic boundary conditions, field values shifted by $2\pi$ must be identified; in conventional language it is said that the field $h$ is compactified on a circle of radius $1$: $h\sim h+2\pi$. 

The dimers are viewed as elementary electric charges, i.e.\ are created by the field $\cos(h)$ (the simplest function of $h$ that satisfies the compactification $h\sim h+2\pi$). The standard computation \cite{Nienhuis} gives for the dimer-dimer correlator decay exponent:
\begin{equation}
 \alpha=\frac{1}{\kappa}\ .
\end{equation}
This yields the stiffness for the dimer model $\kappa_D=1/2$. Elementary magnetic charges can also identified with monomers, and the corresponding exponent describing their two-point function is given by $\beta=\kappa$, so $\beta_D=1/2$.
Once the stiffness fixed to the appropriate value, Eq.~(\ref{eq:free_field_bis}) describes all the universal properties of the classical dimer model. By virtue of the correspondence between correlators in the classical model and those in the dimer RVB state, the same exponents apply to the correlators in the latter as well. 
The same Coulomb-gas techniques can be used on the six-vertex model as well, yielding a value of $\kappa$ continuously varying with the couplings \cite{Nienhuis}.

In this framework the quantum dimer and quantum six-vertex models can also be written in terms of a free scalar field in the 2+1 dimensional ``quantum Lifshitz'' model \cite{Henley,QuantumLifshitz}. This is a field-theory version of an RK Hamiltonian, where the equal-time correlators of the 2d {\em quantum} model are those of a 2d conformal field theory. This model is critical but not Lorentz-invariant, with dynamical critical exponent $z=2$.

Because of the arguments above, it is reasonable to hope that all the equal-time correlators in the $SU(N)$ RVB state for any $N$ can be described by this Coulomb gas, albeit with stiffness depending on $N$. Substantial evidence for this was provided in the $SU(2)$ case in \cite{RVB2}, where several universal quantities, including the dimer-dimer and monomer-monomer decay exponents, were found numerically. With appropriate identification of the operators, each yields a independent value for the stiffness.  The numerics gave approximately the same value of the stiffness $\kappa_{2}\approx .83$ for each, providing strong evidence that the Coulomb gas description applies to correlators of {\em spin singlets} in the $SU(2)$ RVB wave function.

\section{Two-cylinder R\'enyi entropy from partition functions}
\label{sec:shape_general}
In this section, we discuss the shape-dependent subleading
contribution\footnote{Of course, the non-universal leading boundary law
(or ``area law'') contribution is shape-independent. Possible additional
universal logarithmic terms, not encountered in the two-cylinder geometry, may
also be shape-independent.} to the entanglement entropy 
of two-dimensional gapless systems discussed in \cite{Ju2012}. For R\'enyi index $n$ larger than a certain critical value $n_c$, we derive it exactly for the scaling limit of the square-lattice dimer RVB state, the quantum six-vertex state, and the entire quantum critical line of the quantum Lifshitz theory. Thus we expect it to apply to the scaling limit of the $SU(N)$ RVB state as well, with an appropriate parameterization of $\kappa$ in terms of $N$.

In particular, we show that when a  torus is split into two cylinders labeled $A$ and $B$, the R\'enyi entropy describing the entanglement between the two cylinders can be rewritten in terms of conformal field theory partition functions ${\cal Z}$ as:
\begin{equation}\label{eq:shape_general}
 s_{n>n_c}=\frac{n}{1-n}\ln \left(\frac{{\cal Z}_{A}{\cal Z}_B}{{\cal Z}_{\rm torus}}\right)\ .
\end{equation}
Below we detail the precise definition of the CFT partition functions, but the important fact is that for the cases at hand, these are known. Thus this expression gives an explicit formula for this entropy, exact in the scaling limit. The value of $n_c$ depends on the specifics of the model, but $n_c<2$ in all the examples we will consider, so that Eq.~(\ref{eq:shape_general}) applies to the second R\'enyi entropy to be studied numerically in later sections.  Although our derivation applies only to the quantum Lifshitz ground state (i.e.\ RVB-type states that can be written in terms of a two-dimensional classical free boson), we believe it possible that an analogous formula will apply more generally, in particular to any theory with an RK-type Hamiltonian.

\subsection{Entanglement entropy as a Shannon entropy}
\label{sec:eeshannon}

We will now detail how to derive Eq.~(\ref{eq:shape_general}) in the simplest case of quantum dimers on the square lattice. This requires a slight generalization of two ingredients: the mapping to a classical Shannon entropy of Ref.~\cite{Shannonee} (see also \cite{FurukawaMisguich}) and the boundary phase transition argument of Ref.~\cite{Stephan2011} (see Sec.~\ref{sec:bpt}). Although this derivation does not apply to the $SU(N)$ RVB case, we will later explain why we think Eq.~(\ref{eq:shape_general}) should still apply.

The orthogonality of the different dimer configurations in the dimer RVB state allows for huge technical simplifications. As was shown in Ref.~\cite{Shannonee}, the entanglement entropy can be expressed as a classical Shannon entropy. We give here only a brief summary of the results, and refer to \cite{Shannonee} for more details. 

\begin{figure}[ht]
\begin{center}
\includegraphics[scale=0.8]{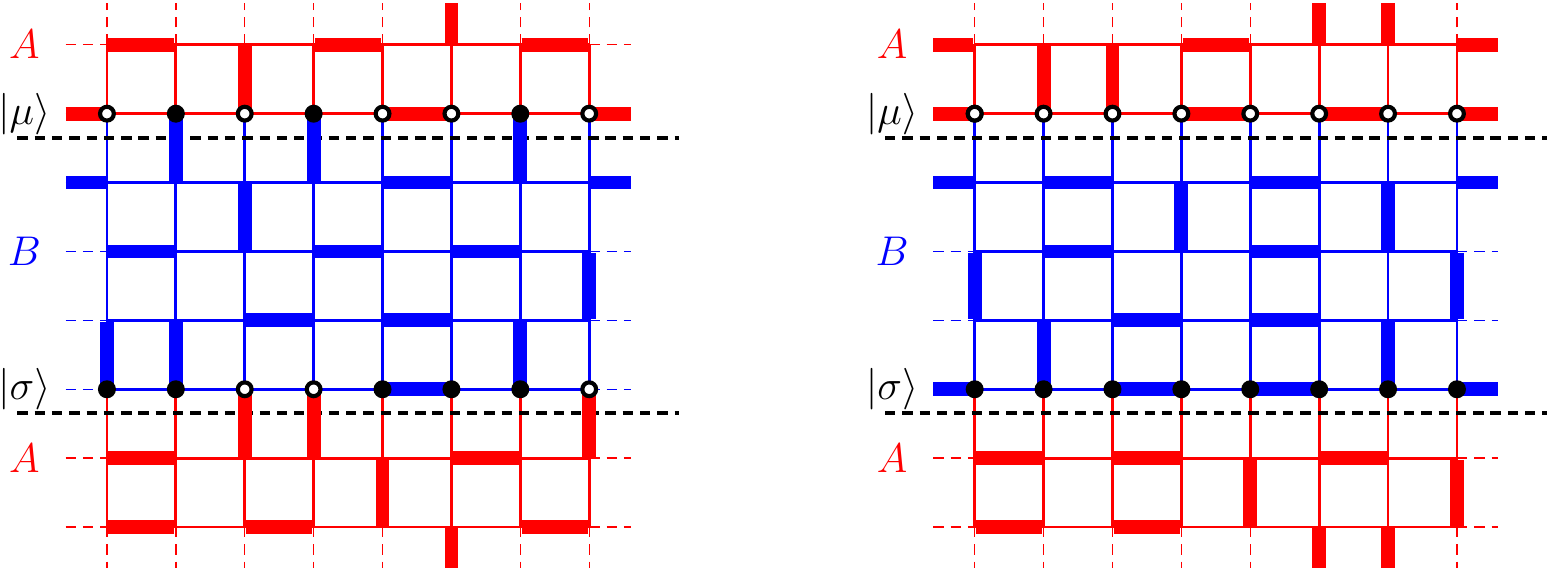}
\end{center}
\caption{Bipartition of the $8\times 8$ torus. The two boundaries between $A$ and $B$ are emphasized by thick dashed lines. Links crossing the lower boundary are defined to be in $A$, whereas those crossing the upper boundary in $B$.}
\label{fig:bipartition}
\end{figure}

Let us cut our torus in two subsystems $A$ and $B$, as is shown for example in Fig.~\ref{fig:bipartition}. We label the dimer configurations along the boundaries as $|\sigma\rangle$ and $|\mu\rangle$.
The main difference with \cite{Shannonee} is that there are two boundaries here, but the the same arguments apply. The von Neumann entropy can be recast as a classical Shannon entropy
\begin{equation}
 S=-\sum_{\sigma,\mu} p_{\sigma,\mu} \ln p_{\sigma,\mu}\ ,
\end{equation}
where the $p_{\sigma,\mu}$ are the probabilities of a given boundary configuration $|\sigma\rangle$,$|\mu\rangle$ between $A$ and $B$. 
\label{sec:lg.}
The $\psm$ are precisely the eigenvalues of the reduced density matrix. Said differently, the Schmidt decomposition of the ground state reads
\begin{equation}\label{eq:schmidt}
|\Psi_D\rangle=\sum_{
\sigma,\mu} \sqrt{p_{\sigma,\mu}}|\psi_{\sigma,\mu}^A\rangle |\psi_{\sigma,\mu}^B\rangle,
\end{equation}
where $|\psi_{\sigma,\mu}^A\rangle$ is a superposition of all dimer states in $A$ compatible with the boundary conditions $\sigma$ and $\mu$, and the same goes for $B$. All these vectors are mutually orthogonal:
\begin{equation}\label{eq:schmidt_orthogonality}
\langle \psi_{\sigma,\mu}^\Omega|\psi_{\sigma^\prime,\mu^\prime}^{\Omega^\prime}\rangle=\delta_{\sigma \sigma^\prime}\delta_{\mu \mu^\prime}\delta_{\Omega \Omega^\prime}\ .
\end{equation}
This mapping therefore allows the R\'enyi entanglement entropy to be rewritten as
\begin{equation}
 S_n=\frac{1}{1-n}\ln \left(\sum_{\sigma,\mu}[\psm]^n\right).
 \label{Spsm}
\end{equation}
We name this quantity the R\'enyi-Shannon entropy in the following. The probabilities can be conveniently rewritten in terms of dimer partition functions as 
\begin{equation}
 \psm=\frac{Z_{\sigma,\mu}}{Z},
 \label{Zpsm}
\end{equation}
where $Z_{\sigma,\mu}$ is the number of dimer configurations compatible with the boundary configuration $|\sigma\rangle,|\mu\rangle$, while $Z$ is the number of dimer coverings on the whole torus. 

It is important to emphasize the main reasons why such a remarkable quantum-classical correspondence holds. While (\ref{eq:schmidt}) remains true for all the states we consider in this paper, the orthogonality (\ref{eq:schmidt_orthogonality}) is guaranteed only if three other conditions are satisfied:
\begin{enumerate}
 \item The orthogonality of quantum states corresponding to different classical configurations. 
 \item A certain type of local constraint \cite{Shannonee}, which for ${\mathbb Z}_2$ variables on the bonds of the square lattice (e.g.\ dimer occupancies) amounts to:  if three of the variables around each site are specified, then the value of the fourth is uniquely determined.
 \item Interactions are only between bond variables sharing a common site.
 \end{enumerate}
Quantum dimers obviously satisfy all these requirements. The same goes for the quantum six-vertex state, (ii) being guaranteed by the ice rule. The $SU(N)$ RVB states satisfy (ii) and (iii) but not (i) (see (\ref{eq:nonortho})). An example of a Coulomb-gas wave function that satisfies (i) and (ii) but not (iii) is a dimer wave function built out of an interacting dimer model \cite{Alet_dimers1,Alet_dimers2}. Let us finally mention that when Eq.~(\ref{eq:schmidt_orthogonality}) does not hold, it is sometimes possible to nevertheless obtain the Schmidt eigenvalues by numerically diagonalizing the matrix whose elements are the $\langle \psi_{\sigma,\mu}^\Omega|\psi_{\sigma^\prime,\mu^\prime}^{\Omega^\prime}\rangle$ \cite{VBS_EE1,VBS_EE2}.  

\begin{figure}[ht]
 \begin{center}
 \includegraphics[width=10cm]{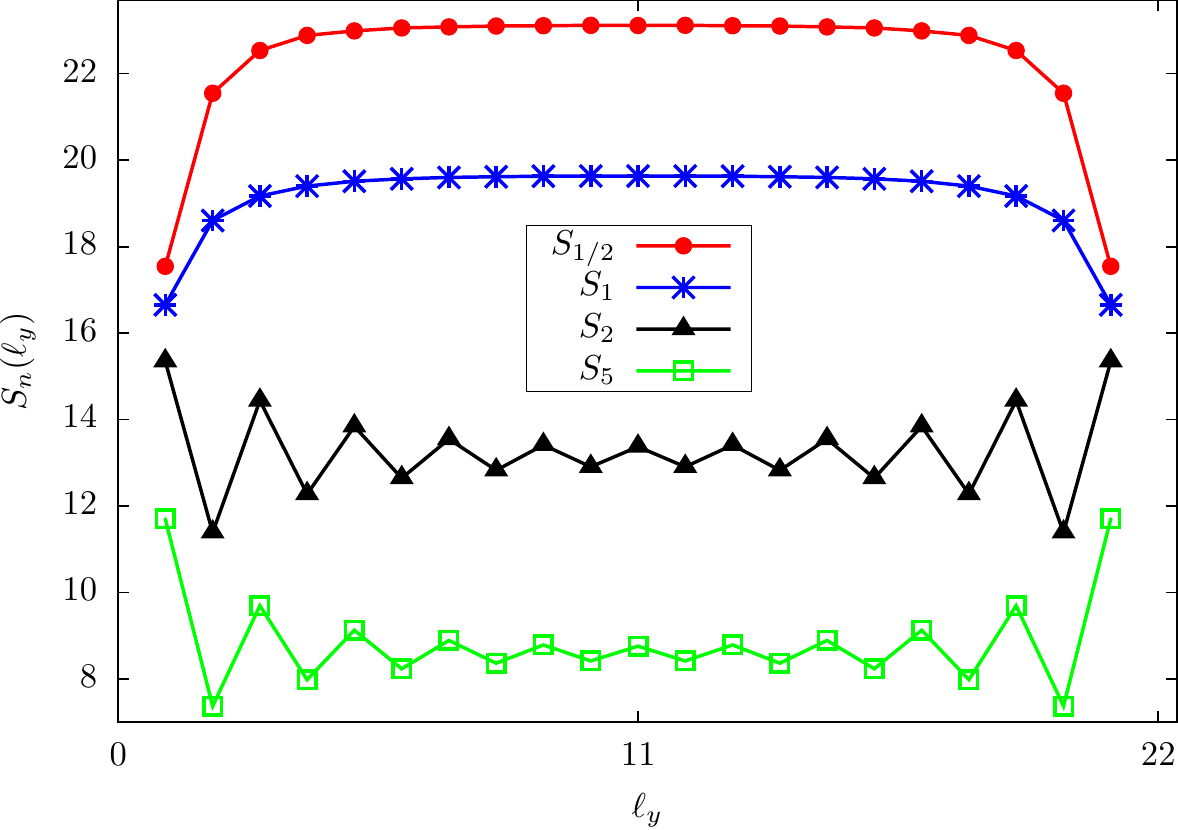}
 \end{center}
\caption{Two-cylinder R\'enyi entanglement entropies $S_{1/2}$, $S_1$, $S_2$ and $S_5$ for the dimer RVB state on the square lattice in the torus geometry. Here $L_x=L_y=22$.}
\label{fig:ee_d_torus}
\end{figure}

Returning to the dimer RVB state, these expressions make it possible to use the computer to find essentially exact values for the R\'enyi entropy. Using free-fermion techniques described in \ref{sec:lgv}, each probability can be expressed as a product of determinants of two matrices of size $\sim L/2$, and computed numerically in a time of order $\approx L^3$. Using translational symmetry along the $x$ axis as well as the conservation of winding numbers, we are able to compute any R\'enyi entropy up to machine precision in a time of order $L^{3/2}\times 4^L$.
The results for a $22\times 22$ torus are summarized in Fig.~\ref{fig:ee_d_torus}, where the two cylinders are of length $\ell_y$ and $L_y-\ell_y$. 
The figure makes it obvious that the R\'enyi entropy exhibits completely different behavior for different values of $n$. In the ``replica'' phase $n \leq 1$, the curves are relatively flat. In the thermodynamic limit we expect them to go to universal functions, which in principle may be computed within boundary CFT, although we will not study this phase here. For $n$ large enough, we observe an even-odd effect like that observed for the $SU(2)$ RVB state in \cite{Ju2012}. We explain the origin of this transition in dimer RVB state in the next subsection.

\subsection{The entropy in the locked phase $n>n_c$}
\label{sec:bpt}
Here we derive the relation of the two-cylinder entanglement entropy of the dimer RVB state to CFT partition functions in 
(\ref{eq:shape_general}) for $n>n_c$. This follows fairly simply from the description of the long-distance behavior of classical dimers in terms of the free-boson action (\ref{eq:free_field_bis}).

When one utilizes an effective field-theory action to describe the scaling limit of a lattice model, one must include all terms in the action consistent with the symmetries of the field theory. As is familiar from the derivation of the Kosterlitz-Thouless transition in field theory, operators of the form 
\begin{equation}
 V_d=-\cos \left(d \,(h-h_0)\right)
 \label{Vdef}
\end{equation}
with $d$ integer 
are consistent with the compactification  of the boson $h\to h + 2\pi$.
If such a term is relevant, then the action flows to a configuration where the height $h$ is locked to one of the minimal values $h=h_0$ mod $2\pi/d$. For dimers on the square lattice, it has long been known that any such term allowed is irrelevant in the bulk; this is why the action (\ref{eq:free_field_bis}) applies and the correlators are algebraically decaying. However, we showed in the preceding subsection \ref{sec:eeshannon} that the probabilities used to compute the R\'enyi entropy depend on the boundary conditions. Thus the crucial observation of \cite{Stephan2011} (see also \cite{Shannonee}) is that one needs to include terms of the form (\ref{Vdef}) localized at the {\em boundary}.

To be precise, in the continuum limit the probabilities $p_{\sigma,\mu}$ depend now on the configurations of the field $p(\phi)$. The probability of observing field configuration $\phi$ at the boundary is still Gaussian
\begin{equation}
 p(\phi)\propto \exp(-S_{\kappa}[\phi]),
\end{equation}
and raising this probability to the $n-$th power yields
\begin{equation}
 [p(\phi)]^n=\exp(-n S_{\kappa}[\phi])=\exp(-S_{n\kappa}[\phi]).
\end{equation}
Therefore, the system near the boundary effectively feels a stiffness $\kappa^\prime=n\kappa$. 
Standard Coulomb-gas/field-theory computations (see e.g.\ \cite{Ginsparg,Nienhuis}) then give the boundary scaling dimension of (\ref{Vdef}) to be $d^2/(2n\kappa)$. 
As a consequence, $V_d$ is irrelevant as long as $d^2>2n\kappa$. Thinking of $n$ as a continuous parameter, this defines two regions, separated by a (boundary) Kosterlitz-Thouless phase transition. The critical value of $n$ in the computation of  $S_n$  is therefore
\begin{equation}
 n_c=\frac{d_{\rm min}^2}{2\kappa}\ ,
\end{equation}
where $d_{\min}$ is the smallest $d$ allowed by the lattice symmetries. For dimers on the square lattice $d_{\rm min}=1$ and $\kappa=1/2$, so that the critical value is
\begin{equation}
 n_c=1.
\end{equation}
The two qualitatively different types of behavior seen in fig.\ \ref{fig:ee_d_torus} indeed occur on opposite sides of $n_c$. It is important to note that even though $S_n$ should take a universal form in both phases, the value itself of the critical R\'enyi parameter depends on a degeneracy $d_{\rm min}$, which is non-universal. For example, $d_{\rm min}=2$ for the quantum six-vertex wave function, and $d_{\rm min}=3$ for dimers on the hexagonal lattice. 

This observation allows us to derive the expression for $S_n$ valid for $n>n_c$. In this phase, the boundary operator is relevant. This means that here the boundary value of the field locks on to a minimum of $V_1$. This minimum must be fixed along the entire boundary; this is commonly known as Dirichlet boundary conditions on the boson. We refer therefore to the phase with $n>n_c$ as the ``locked'' phase.

A constant value of the field along the boundary means that the action is at its minimum there, so this is the {\em maximum} value of the probability $p_{\rm max}$. 
Universal contributions to the R\'enyi entropy coming from the sum (\ref{Spsm})  are therefore dominated by the configuration:
\begin{equation}\label{eq:aftert}
 S_{n>n_c}\sim \frac{n}{1-n}\ln (p_{\rm max})
\end{equation}
This result implies that in the scaling limit there is an ``entanglement gap'' between $p_{max}$ and the other values. We have confirmed directly the presence of the entanglement gap for the dimer RVB state by numerical evaluation of the probabilities using the free-fermion techniques detailed in \ref{sec:lgv}.  This is distinct from the quantum Hall case \cite{HaldaneLi,Bernevig_spectrum,Katsura,Estienne,DubailReadRezayi}, where there is no entanglement gap, in correspondence with the gapless edge excitations there.
 
 In terms of dimers, the configuration with maximum probability corresponds to no dimers crossing the boundaries between $A$ and $B$ (see Fig.~\ref{fig:bipartition}). This is a huge technical simplification, since the probability of having no dimers across the cut factorizes into pieces coming from region $A$ and from region $B$. Namely, define $Z_{\rm cyl}(L_x, L_y)$ to be the partition function (the number of dimer configurations) for a cylindrical region of size $L_x$ in the periodic direction and $L_y$ in the other, with boundary conditions corresponding to no dimers leaving the region. Likewise, $Z_{\rm torus}(L_x,L_y)$  is the number of dimer configurations with periodic boundary conditions. Then it follows from (\ref{Zpsm}) that when an $L_x\times L_y$ torus is split into cylinders of lengths $\ell_y$ and $L_y-\ell_y$,
\begin{equation}\label{eq:pmax}
 p_{\rm max}=\frac{Z_{\rm cyl}(L_x,\ell_y)Z_{\rm cyl}(L_x,L_y-\ell_y)}{Z_{\rm torus}(L_x,L_y)} \ .
\end{equation}
The analogous formula still holds if space is a cylinder instead of a a torus, simply by replacing the denominator with the appropriate cylinder partition function. 
Note that the non-universal bulk parts of the partition functions cancel in this
expression. The non-universal boundary parts do not, and contribute to the
 boundary law in (\ref{eq:aftert}). Thus putting (\ref{eq:pmax}) together with
(\ref{eq:aftert}) and filtering out the boundary law term gives
(\ref{eq:shape_general}).

For the dimer RVB state, the partition functions on the cylinder can be computed exactly both on the lattice and in the continuum.  The lattice dimer result is detailed in \ref{sec:dimers_exact}. For the free-boson field theory, these partition functions are known explicitly \cite{BigYellowBook};  we will discuss their application to the two-cylinder R\'enyi entropy in the next section \ref{sec:dimers_entanglement}. In particular, we will show how the even-odd effect is a signature of this locked phase  $n>n_c$, arising from the different boundary conditions necessary for the even and odd sectors.

It should also be possible to extend these results to the region $n<n_c$, using boundary CFT methods \cite{Shannonee,Oshikawa,Zaletel,Stephan2011}. Here, however, the boundary conditions will not allow the factorization of the probability into pieces coming from the two regions. Oshikawa\cite{Oshikawa} has already treated the $y=\ell_y/L_y=1/2$ case for the closely related cylinder geometry, using the replica approach $n\in \mathbb{N}$. There will possibly be subtleties involving analytic continuation in $n$, similar to the two-interval 1d calculation \cite{CCT1,CCT2}. The study of the ``marginal'' case $n=n_c$ would presumably be even more challenging \cite{Stephan2010,Stephan2011}.

\section{Entanglement in the dimer RVB state}
\label{sec:dimers_entanglement}

This section is devoted to the exploration of the consequences of Eq.~(\ref{eq:shape_general}) for the two-cylinder R\'enyi entropy of the dimer RVB state. We will calculate explicitly the partition functions, and so find results exact in the scaling limit. 

The main subtlety in this computation is how to account in the field theory for the distinct results occurring when the cylinders are of even and of odd length, as apparent in the numerical results in fig.\ \ref{fig:ee_d_torus} for the dimer RVB state and those in \cite{Ju2012} for the $SU(2)$ RVB state. The reason this is possible in the field theory is that a fixed/Dirichlet boundary condition is in fact a family of boundary conditions, corresponding to the specific value $h_0$ that the field takes at a boundary. This value can be always be shifted overall, since the bulk action used to compute the partition functions is independent of it. However, on a cylinder, there are two separate boundaries, and what cannot be shifted away is the {\em difference} of the values on the two ends \cite{OshikawaAffleck1,OshikawaAffleck2}.    
The general arguments given above in section (\ref{sec:bpt}) do not specify this difference; it follows rather from the particular microscopic model. For dimers, we showed in section \ref{sec:cg} that 
\begin{equation}
 \Delta h =2\pi (w+a)\quad,\quad w \in \mathbb{Z},
\end{equation}
where $a=0$ for even $L_y$ and $a=1/2$ for odd $L_y$.

\begin{figure}[t]
\begin{center}
 \includegraphics[width=8cm]{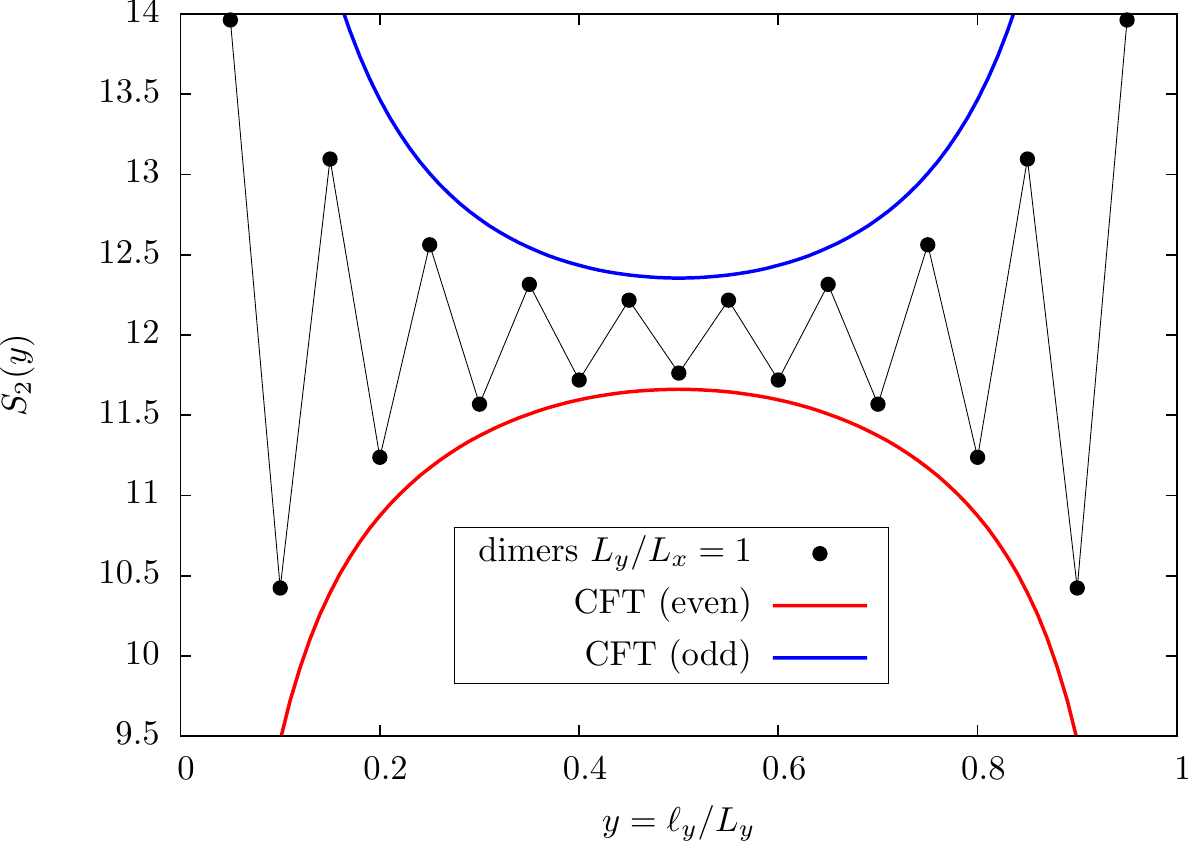}
 \includegraphics[width=8cm]{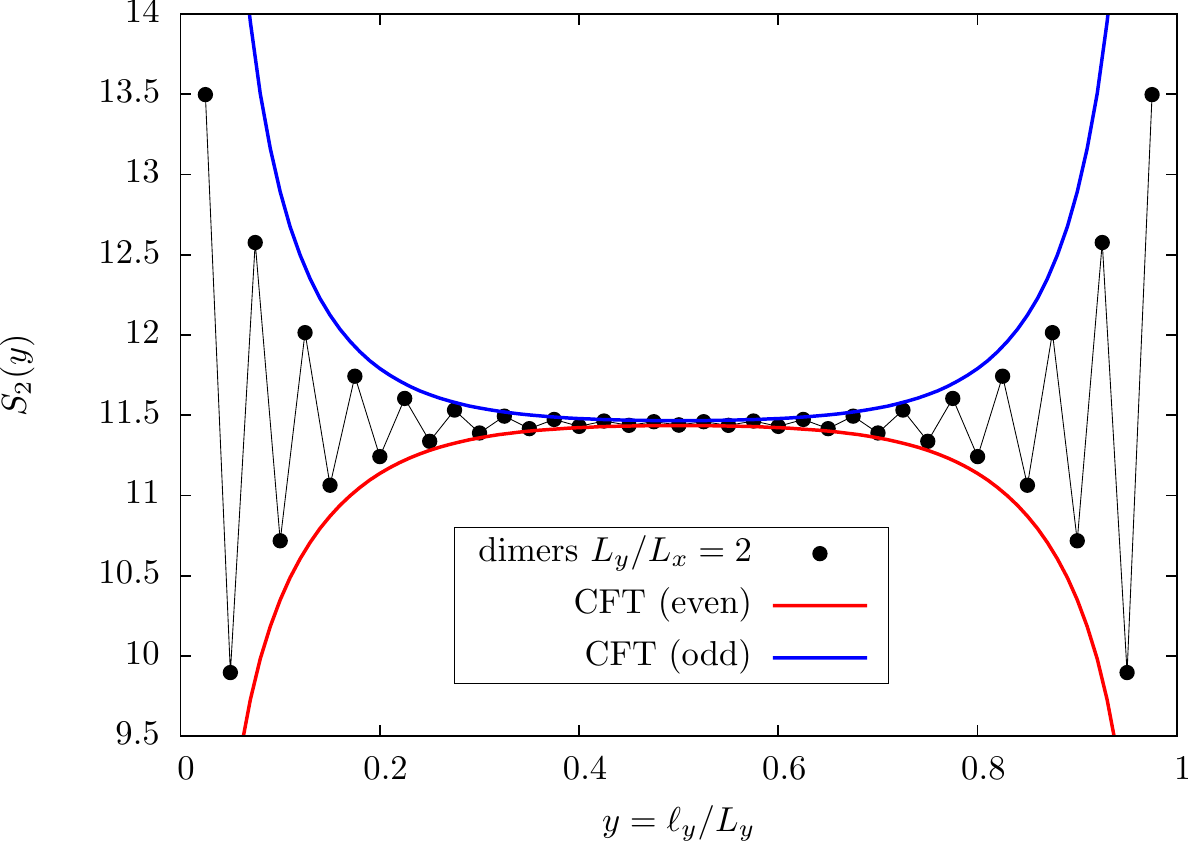}
 \end{center}
 \caption{Numerical results for the second R\'enyi entropy $S_2$ as a function of the subsystem-ratio $y=\ell_y/L_y$ for $L_x=20$. Black dots are the data for square-lattice dimer RVB state, while the CFT predictions are given by Eq.~(\ref{eq:cft_prediction}) in the even case and Eq.~(\ref{eq:cft_prediction_odd}) in the odd case. \emph{Left:} Torus with aspect ratio $L_y/L_x=1$. \emph{Right:} Torus with aspect ratio $L_y/L_x=2$.}
 \label{fig:dimers_shape}
\end{figure}

The computation of the free-boson partition function is completely standard; see e.g.\ \cite{EggertAffleck,FSW,BigYellowBook}. Since the action is quadratic in the bosonic field, it can be decoupled into an oscillator part and a classical part. Only the classical part is affected by the compactification $h\sim h+2\pi$, and by the boundary conditions. The partition function for Dirichlet boundary conditions on both ends of the cylinder, with the height differing by $2\pi a\, {\rm mod }\,2\pi$, is given by
\begin{equation}
 \mathcal{Z}_{\rm cyl}^{DD^{(a)}}=q^{-1/24}\prod_{w=1}^{\infty}\left(1-q^w\right)^{-1}\sum_{w=-\infty}^{+\infty}q^{\,\kappa (w+a)^2}\quad,\quad q=e^{-\pi L_x/L_y}
\end{equation}
It is convenient to rewrite such sums in terms of the Jacobi theta and Dedekind eta functions defined in \ref{sec:CFT_Jacobi}, to take advantage of many elegant identities. 
The even case $a=0$ can be written, using (\ref{eq:eta_def},\ref{eq:theta3_def},\ref{eq:thetas},\ref{eq:modtheta3},\ref{eq:modeta}), as
\begin{equation}\label{eq:cylinder_dd}
 \mathcal{Z}_{\rm cyl}^{DD}(\tau)=\frac{\theta_3(2\tau)}{\eta(2\tau)}\ ,
\end{equation}
where we adopt the conventional variable  $\tau=i L_y/L_x$, and have set $\kappa=1/2$, the dimer value.
For the odd case $a=1/2$, we have, using this time (\ref{eq:eta_def},\ref{eq:theta2_def},\ref{eq:thetas},\ref{eq:modtheta4},\ref{eq:modeta}):
\begin{equation}
\label{eq:cylinder_ddprime}
 \mathcal{Z}_{\rm cyl}^{DD^\prime}(\tau)=\frac{\theta_4(2\tau)}{\eta(2\tau)}\ .
\end{equation}
Plugging these results into Eq.~(\ref{eq:shape_general}) and using the torus partition function \cite{Ginsparg,BigYellowBook} gives for the two-cylinder entanglement entropies
\begin{eqnarray}\label{eq:cft_prediction}
 s_n^{\rm (even)}(y,\tau)&=&\frac{n}{1-n}\ln \left(\frac{\eta(\tau)^2}{\theta_3(2\tau)\theta_3(\tau/2)}\times\frac{\theta_3(2y \tau)\theta_3(2(1-y)\tau)}{\eta(2y\tau)\eta(2(1-y)\tau)}\right)\ ;\\
 \label{eq:cft_prediction_odd}
 s_n^{\rm (odd)}(y,\tau)&=&\frac{n}{1-n}\ln \left(\frac{\eta(\tau)^2}{\theta_3(2\tau)\theta_3(\tau/2)}\times\frac{\theta_4(2y \tau)\theta_4(2(1-y)\tau)}{\eta(2y\tau)\eta(2(1-y)\tau)}\right)\ .
\end{eqnarray}
Fig.~\ref{fig:dimers_shape} shows a numerical test of the universal shape Eq.~(\ref{eq:cft_prediction}) for the dimer RVB states with two different aspect ratios $L_y/L_x=1$ and $L_y/L_x=2$.  
Obviously, these $n>n_c$ results are quite different in the even and odd sectors. The difference is proportional to $n/(n-1)$, and therefore slightly diminishes with increasing $n$, but does not go away.

This even/odd effect disappears as $L_y/L_x$ is increased; it follows from the CFT expressions that this occurs exponentially quickly.  It does not approach the celebrated 1d result\cite{Cardy} $\propto \ln\sin (\pi y)$ in the effective 1d limit $L_y/L_x\to\infty$, because the RK Hamiltonian becomes gapped with correlation length $\xi\propto L_x$ in this limit.

\begin{figure}[ht]
 \begin{center}
  \includegraphics[width=8cm]{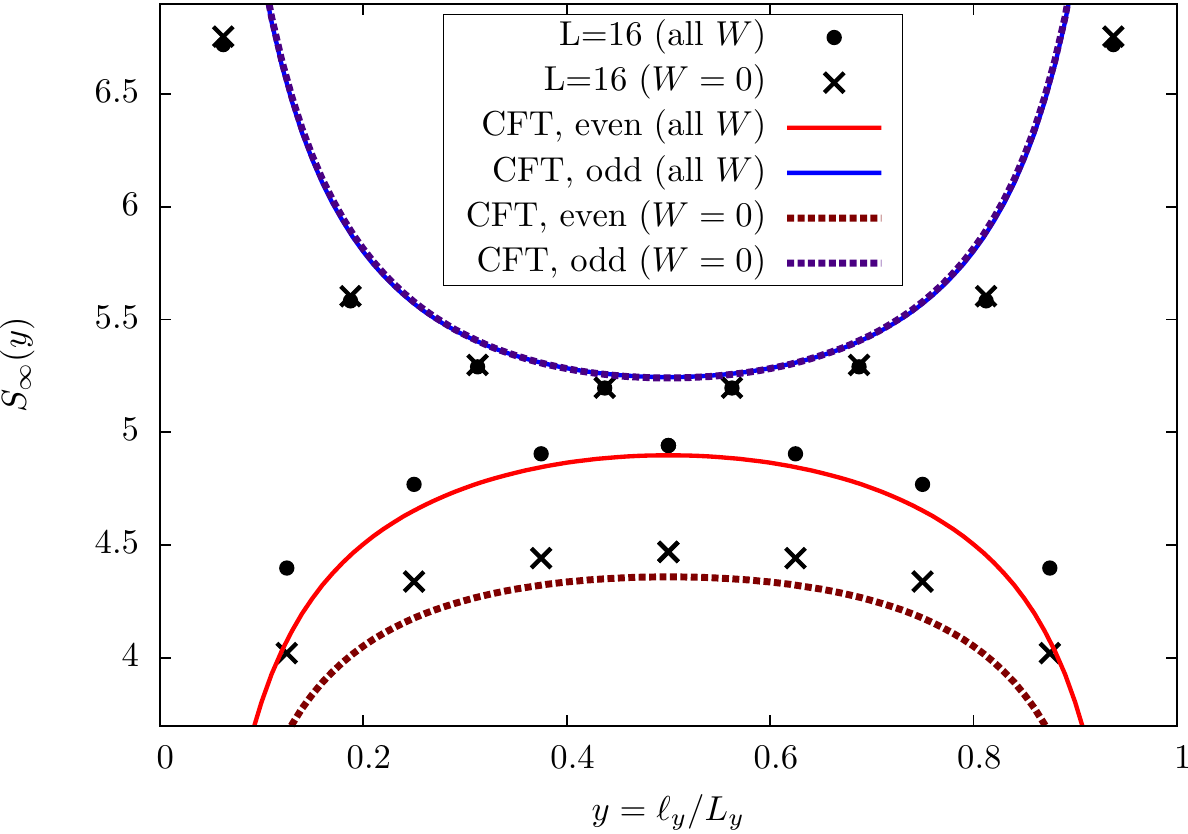}
  \includegraphics[width=8cm]{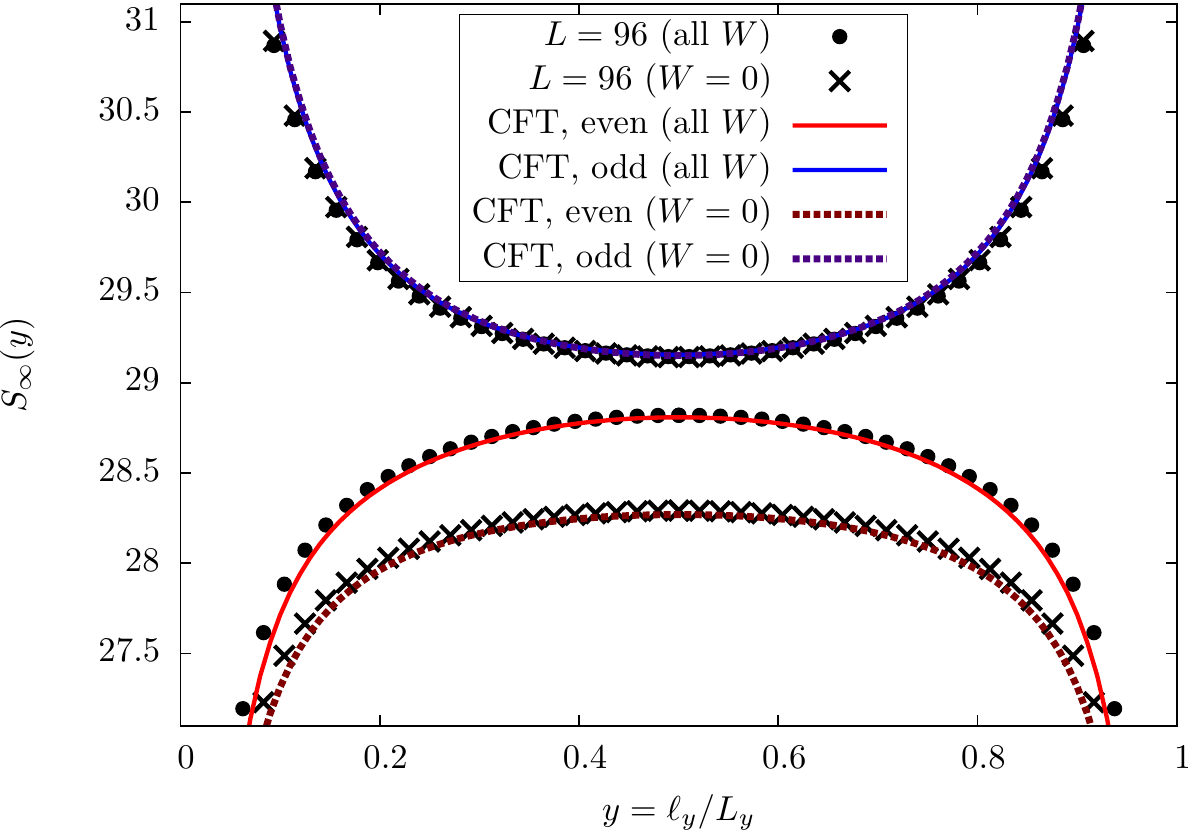}
 \end{center}
\caption{Comparison of numerics and analytics in the exactly solvable case $S_\infty$. Black points are the numerical data for all winding sectors, and black crosses represents the numerical data in the $W=(0,0)$ sector. \emph{Left:} $L_x=16$, the data is compared with Eqs.~(\ref{eq:cft_prediction}), (\ref{eq:cft_prediction_odd}),  (\ref{eq:cft_w0}) and (\ref{eq:cft_w0_odd}) up to a single constant. \emph{Right:} Same procedure for $L_x=96$. }
\label{fig:Sinfty}
\end{figure}
The relatively large finite-size effects apparent in the plots should disappear in the thermodynamic limit.  To illustrate the slowness of the convergence, we study the exactly solvable case $n\to\infty$, where the agreement between CFT and the lattice can be made rigorous. 
It follows from (\ref{Spsm}) that only the largest probability (\ref{eq:pmax}) contributes: $S_\infty=-\ln p_{\rm max}$, and so numerical analysis is much easier. We checked system sizes up to $L_x=704$, and found that while indeed there are strong finite-size effects, the data converges in the end to the CFT result. 
The slow convergence is illustrated by the numerical and analytical results in performed in Fig.~\ref{fig:Sinfty}, for the aspect ratio $L_y/L_x=1$ and the two system sizes $L_x=16$ and $L_x=96$. 
We also computed $s_\infty$ in the $W=(0,0)$ winding sector directly for the dimer model using the techniques described in the appendices, finding
\begin{eqnarray}\label{eq:cft_w0}
 s_{\infty,0}^{\rm (even)}&=&-\ln \left(\frac{\eta(\tau)^2\theta_3(2y(1-y)\tau)}{\eta(2y\tau)\eta(2(1-y)\tau)}\right);\\\label{eq:cft_w0_odd}
 s_{\infty,0}^{\rm (odd)}&=&-\ln \left(\frac{\eta(2\tau)^2\theta_4(2y(1-y)\tau)}{\eta(2y\tau)\eta(2(1-y)\tau)}\right).
\end{eqnarray}
Even though the winding numbers are zero, the result still depends on the compactification radius because of winding fluctuations at the boundaries. 
Interestingly, the even-odd effect is enhanced in the $W=(0,0)$ winding sector, in agreement with CFT. We also observe that the odd curves in both sectors almost coincide on the lattice, as well as in the scaling limit. This coincidence can be explained from the CFT result at $y=\ell_y/L_y=1/2$:
\begin{equation}
 s_{\infty}^{\rm (odd)}(y=1/2)-s_{\infty,0}^{\rm (odd)}(y=1/2)=2q^2+\mathcal{O}(q^4) \ ,\qquad q=e^{-\pi}.
\end{equation}
This is a consequence of cancellations at the free fermion point $\alpha=2$; in general $$
 s_{\infty}^{\rm (odd)}(y=1/2)-s_{\infty,0}^{\rm (odd)}(y=1/2)=2(q^{1/\alpha}-q^{\alpha/4})-2(q^{2/\alpha}-q^{\alpha/2})+(8/3)(q^{3/\alpha}-q^{3\alpha/4})+\mathcal{O}(q^{\alpha})\ .$$

\section{Correlations in the $SU(N)$ RVB wave function}
\label{sec:correlations}
We now start to explore the question of universality, with the motivation of understanding whether the results obtained for the dimer RVB state can be adapted to describe the $SU(N)$ RVB states. In this section we use quantum Monte Carlo techniques to find the spin-spin and dimer-dimer correlators for $N=2,3,4,5$, generalizing some of the results of \cite{RVB1,RVB2}. This shows that there is no evidence for a transition as $N$ is varied, a strong piece of evidence for universality. It also allows us to make contact with the results of \cite{Damle}, where an expansion around the dimer RVB state is developed.

\subsection{Monte Carlo methods}
\label{sec:numerics}

To study $SU(N)$-RVB wavefunctions, we use an unbiased Monte Carlo technique that computes expectation values in an equal-amplitude superposition of states using a Metropolis importance sampling scheme \cite{RVB2}.  Numerically, the RVB wavefunction is represented in a combined VB-spin (or dimer-spin) basis, $|V_{\mathcal C} \rangle = |D_{\mathcal C} \rangle |Z_{\mathcal C} \rangle$, where $| D_{\mathcal C} \rangle$ is a list of site-pairs, $[i,j]$, specifying the ends of the valence bonds, and $|Z_{\mathcal C} \rangle$ is the spin state for each site on the lattice.

In order to obtain expectation values (as described below), two RVB wavefunctions are sampled simultaneously, $\langle V_{\mathcal C} |$ and $| V_{\mathcal C'} \rangle$.  The two VB states $\langle D_{\mathcal C} |$ and $| D_{\mathcal C}' \rangle$ can initially be chosen at random; spin states compatible with a non-zero overlap $\langle V_{\mathcal C} | V_{\mathcal C'} \rangle$ are then constructed.  Since the spin basis is orthogonal, we set $|Z_{\mathcal C} \rangle = | Z_{\mathcal C}' \rangle$ to ensure a non-zero overlap.  Then, the Monte Carlo sampling scheme proceeds in two distinct steps.   

After initializing the wavefunction in one configuration $|D_{\mathcal C}  \rangle | Z_{\mathcal C}  \rangle$, the first step is to sample a new spin state $|Z_{\mathcal C}   \rangle = |Z_{\mathcal C} ' \rangle$.  This is done at random, subject to the condition that the overlap $\langle V_{\mathcal C} | V_{\mathcal C'} \rangle$ remains non-zero.  The second step is to sample the VB states $|D_{\mathcal C}  \rangle$ and $|D_{\mathcal C} ' \rangle$.  This must be done subject to the constraint imposed by the spin states.
As in the $SU(2)$ case \cite{RVB2,Ju2012}, one has the choice of ``local'' or ``non-local'' updating schemes to sample all possible VB states \cite{sandvik2010loop}.
\begin{figure}[t]
 \begin{center}
    \scalebox{.3}{\includegraphics[width=\columnwidth]{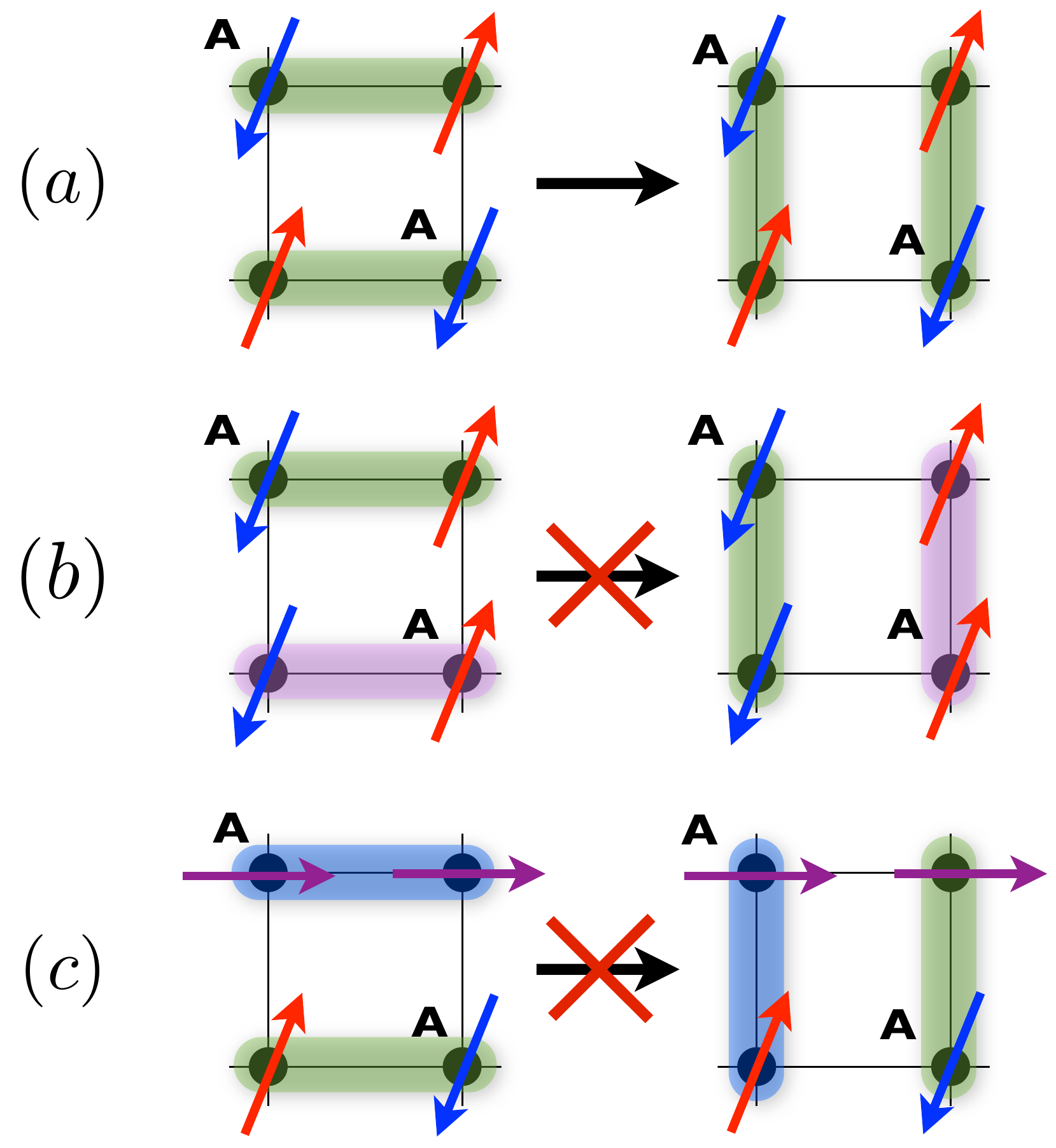}}
 \end{center}
 \caption{Possible local updates of the valence-bond (VB) configuration $|D_{\mathcal{C}} \rangle$ in a Monte Carlo simulation, where thick lines are the VB states.  Figures a) and b) show cases for $SU(2)$, where in a) the underlying spin configuration ($m_i = \pm \frac{1}{2}$) allows the update shown, while in b) it is precluded.  In c), an $SU(3)$ case is shown with an additional spin state $m_i = 0$ (top two sites), which precludes the update shown.}
 \label{fig:updates}
 \end{figure}
In a local update, the simulation chooses two parallel valence bonds on a square plaquette and moves each to the empty bond, subject to the condition that the spin-basis is compatible with this new configuration: Fig.~\ref{fig:updates}(a).
However, if the spin-basis is not compatible with this proposed change, as in Fig.~\ref{fig:updates}(b-c) where the right pointing arrow represents a different flavor spin for an $SU(N>2)$ system (e.g.\ $m_i=0$ for $SU(3)$), the move is rejected, and a new ``plaquette flip" is proposed by the simulation.
While simple to perform, this move is not ergodic in that it cannot change the winding number of the system.
Therefore, this update is most useful if one wants to fix the winding number to do measurements (e.g.\ for the correlation measurements below, we use the winding number $W=(0,0)$).
In contrast, a non-local loop update moves through the ensemble of states by creating a defect at some lattice site and propagating it through the system (subject to the spin constraints) until the defect reaches its initial point, thus closing the loop \cite{sandvik2010loop}.

The important point to note is that the Monte Carlo update for the RVB wavefunction proceeds in two steps: sampling the spin state $|Z_{\mathcal C} \rangle$, and sampling the VB states $|D_{\mathcal C} \rangle$.  Each basis configuration is constrained by the other such that the valence-bond remains a quantum $SU(N)$ singlet.  In contrast, for the classical dimer model, no spin state exists, and dimer wavefunctions are orthogonal.

\begin{figure}[t]
 \begin{center}
    \scalebox{.6}{\includegraphics[width=\columnwidth]{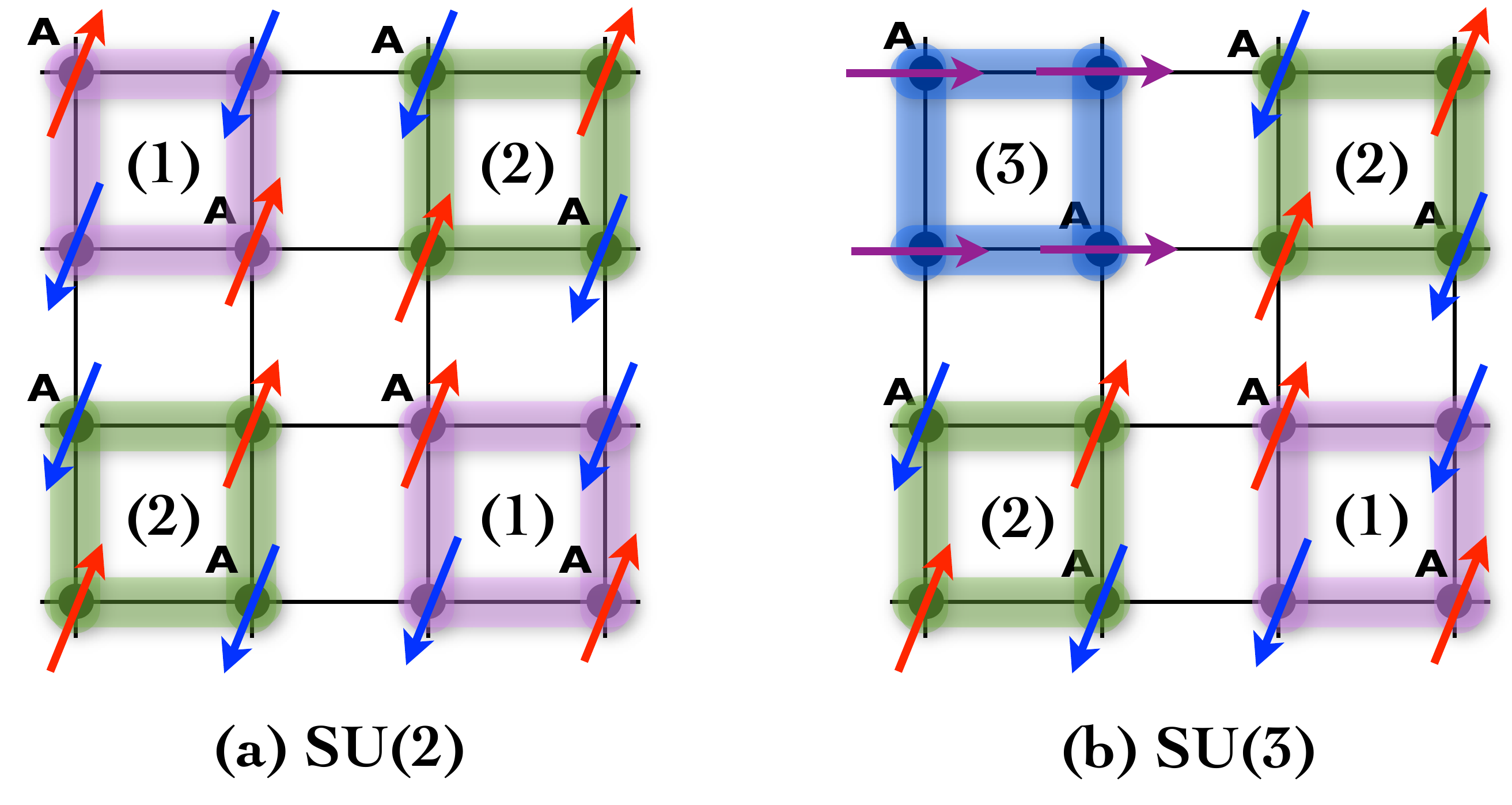}}
 \end{center}
 \caption{Simple transition graphs for RVB wavefunctions.  In a), an $SU(2)$ state illustrates two different types of transition graph, with ``loops'' labeled 1 and 2, based on whether a spin $m_i$ is $+\frac{1}{2}$ or $-\frac{1}{2}$ on sublattice A.  VB updates must occur between loops of the same type (as in Fig.~\ref{fig:updates}a)).  In b), an $SU(3)$ state has three loop types, depending on whether $m_i$ is +1, -1 or 0 on sublattice A.
 \label{fig:su2su3graphs} }
 \end{figure}

We now discuss the methods we have used to generalize the $SU(2)$ updates used in Ref.~\cite{RVB2,Ju2012} to account for the complexities introduced by the $SU(N)$ RVB wavefunctions.
We do this by considering the similarities and the differences between the $SU(2)$ and $SU(3)$ wavefunctions.
Given Eq.\ (\ref{eq:rvb}), the singlets on a pair of sites $[i,j]$ can be written as
\begin{eqnarray}
\left[i.j\right]_{N=2} &=& \frac{1}{\sqrt{2}} \big[ | \!\uparrow \rangle_i | \!\downarrow \rangle_j - | \!\downarrow \rangle_i |\! \uparrow \rangle_j \big] , \quad \uparrow,\downarrow = \pm \frac{1}{2}\\
\left[i.j\right]_{N=3} &=& \frac{1}{\sqrt{3}} \big[ |\! \uparrow \rangle_i | \!\downarrow \rangle_j + |\! \downarrow \rangle_i |\! \uparrow \rangle_j - |0\rangle_i |0\rangle_j \big] , \quad \uparrow,\downarrow = \pm 1.
\label{eq:su2su3}
\end{eqnarray}
While sampling the spins, there are $N$ different spin flavors to choose for a site on sublattice A.
Once this spin is chosen, all other spins in the same loop are determined by Eq.~\ref{eq:rvb}.
For instance, consider the $SU(2)$ wavefunction in Fig.~\ref{fig:su2su3graphs}(a).
There are \textit{two} types of loops: one with $\uparrow$ (loop 1) and another with $\downarrow$ (loop 2) on sublattice A.
Once the spin basis is chosen in the transition graphs, the VB configurations in the bra and ket are independently updated as prescribed above according to the same spin basis.
Updates, however, are only possible between loops of the same type (1 or 2).
Therefore, instead of keeping track of the spins on each site, it suffices to keep track of the loop labels at each site.

Then, generalizing this a little further, consider Fig.~\ref{fig:su2su3graphs}(b).
One can easily see that there are three loop types: $\uparrow$, $\downarrow$, and 0 on sublattice A.
So, for an $SU(N)$ system, there are $N$ different loop types, and we quickly run into ergodicity issues as updates are more readily rejected.
We find that as $N>4$, the number of updates that one can successfully make becomes much smaller, making the statistical error bars much slower to converge in Monte Carlo.
As shown by our data, some estimators for this large $N$ show signs of ergodicity loss; this might be cured by a more sophisticated sampling scheme.

In our VB basis Monte Carlo simulations, we begin by computing the spin-spin and the dimer-dimer correlations. While the two-point spin correlation is well-known~\cite{sutherland1988monte,liang1988some}, the four-point spin interaction was recently derived in~\cite{beach2006some}. The expectation values were then generalized in~\cite{beach2009n} for $SU(N)$ spins. Using the notation $(ij)_L$ for two sites $i,j$ in the same loop in the transition graph and $(i)_L(j)_L$ for sites not in the same loop~\cite{RVB2},
\begin{equation}
	C_{\rm spin}({\bf r}_i, {\bf r}_j) 
		= \frac{\langle V_\mathcal{C} | \mathbf{S}_i \cdot \mathbf{S}_j | V_{\mathcal{C}'} \rangle}
			{\langle V_\mathcal{C} | V_{\mathcal{C}'} \rangle}
		 = \left\{ \begin{array}{ccc}
			0 & (i)_L (j)_L\\
			\epsilon_{ij} \, S (S+1) & (ij)_L   \end{array}. \right.
\end{equation}
Here, $\epsilon_{ij} = 1$ if $(i,j)$ are on the same sublattice and $\epsilon_{ij} = -1$ if they are on different sublattices. 
We also measure \textit{parallel} nearest neighbor dimer-dimer correlations, where four sites $i,k,j,l$ are related by $k = i + \mathbf{e}_\alpha, \, l = j + \mathbf{e}_\alpha$ and $j = i + m \mathbf{e}_\beta$, where $\alpha,\beta = x, y, \, \alpha \neq \beta$, and $m \in \{0,1,...\}$.
This simplifies the estimator, which is as follows
\begin{eqnarray}\fl
	C_{dd}({\bf r}_i, {\bf r}_j)  &=& 
			\frac{\langle V_\mathcal{C} | \left( \mathbf{S}_i \cdot \mathbf{S}_{i+\mathbf{e}_\alpha} \right) \left( \mathbf{S}_j \cdot \mathbf{S}_{j+\mathbf{e}_\alpha}  \right) 
					| V_{\mathcal{C}'} \rangle}{\langle V_\mathcal{C} | V_{\mathcal{C}'} \rangle} \\\fl
					 &=& \left\{ \begin{array}{cccc}
				\frac{1}{3} S^2 (S+1)^2 &,& (ij)_L ((i+\mathbf{e}_\alpha)(j+\mathbf{e}_\alpha))_L \;\;{\rm or}\;\; \,(i(j+\mathbf{e}_\alpha))_L ((i+\mathbf{e}_\alpha)j)_L\\
		S^2 (S+1)^2 &,& (i(i+\mathbf{e}_\alpha))_L (j(j+\mathbf{e}_\alpha))_L \end{array}, \right.
\end{eqnarray}
and $0$ otherwise.

In the next section we present data for these correlation functions for $SU(N)$ with $N=2,3,4,5$,  and fit the putative universal exponents associated with their decay.

\subsection{Equal-time correlators}
\label{sec:dimerdimer}

\begin{figure}[ht]
 \begin{center}
  \includegraphics[scale=0.8]{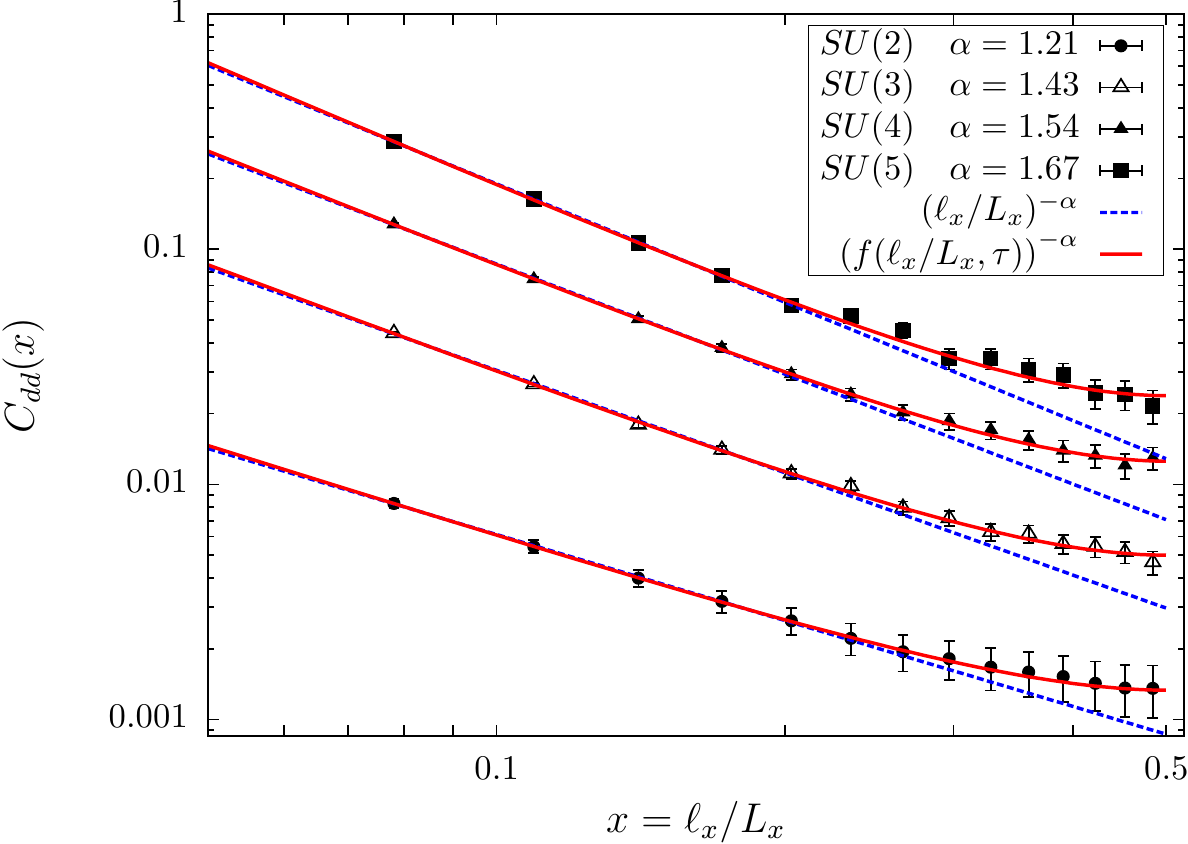}
 \end{center}
\caption{Dimer-Dimer correlations on the $64 \times 64$ torus for $N=2,3,4,5$. Dashed blue lines is a linear fit to Eq.~(\ref{eq:decay}), in the ``linear regime'' $\ell_x \simeq 5,\ldots,15$, while the red solid lines are a fit to the conformal scaling (Eq.~(\ref{eq:cft_fit})) in the same region.}
\label{fig:corr_su2_dimers}
\end{figure}

The connected dimer-dimer correlation function has been recently shown to decay algebraically in the $SU(2)$ case in Refs.~\cite{RVB1} and \cite{RVB2}:
\begin{equation}\label{eq:decay}
 C_{dd}(\mathbf{r}_1,\mathbf{r}_2)\sim \frac{1}{\left|\mathbf{r}_1-\mathbf{r}_2\right|^\alpha}
\end{equation}
at distances large compared to the lattice spacing, but small compared to the torus size $L=L_x=L_y$. Using several different methods, the exponent was found to be approximately $\alpha \simeq 1.2$, independent of the direction.
We extend some of these results to the $SU(N)$ case, mainly focusing on the correlation between two parallel dimers in the transverse direction ($\mathbf{r}_1-\mathbf{r}_2=\ell_x \mathbf{e}_x$). 
Our quantum Monte Carlo results are shown in Fig.~\ref{fig:corr_su2_dimers} for a $L\times L$ torus, with $L=64$. The data clearly indicate that the algebraic decay remains for $N=2,3,4,5$, with $\alpha$ increasing with $N$. 
Since $SU(N)$ dimer correlations reduce to pure dimer correlations in the limit $N\to\infty$, we know that
$
 \lim_{N \to \infty} \alpha_N=\alpha_{\rm D}=2,
$
 and our data is consistent with the smooth interpolation between $N=2$ and $\infty$. 
For each $N$ we extract the leading exponent by a fit (solid blue lines in Fig.~\ref{fig:corr_su2_dimers}) to Eq.~(\ref{eq:decay}) in the regime $1\ll \ell_x \ll L_x$ and for odd $\ell_x$. We find $\alpha_2=1.21(7)$, $\alpha_3=1.43(7)$, $\alpha_4=1.54(8)$, $\alpha_5=1.67(9)$. Notice that as $N$ gets larger, it becomes more and more difficult to converge the Monte-Carlo data in a finite time, resulting in a loss of precision on the corresponding exponent. 

We also observe the presence of a dipolar term which behaves as $(-1)^{\ell_x} \ell_x^{-2}$ for all $N$. This direction-dependent contribution is well known in the dimer model \cite{Alet_dimers2}, and also appears for the $SU(2)$ RVB state \cite{RVB2}. In the dimer model and for the transverse correlations considered here, it cancels exactly the universal term (\ref{eq:decay}) at even distances, so that the dimer-dimer correlation behaves as $\ell_x^{-4}$ \cite{FisherStephenson}. In the $SU(N)$ RVB case the exponent $\alpha$ is different and this does not happen. The dipolar is then a subleading term, which introduces finite-size effect when trying to extract the exponents. For example we get $\alpha_2\simeq 1.14$ at even distances, which is slightly smaller than the previous computed value $\simeq 1.20$ obtained for larger system sizes \cite{RVB1,RVB2}.

These values are in good agreement with the expansion around the dimer RVB state developed in \cite{Damle}. This is a cluster expansion of the loop model, describing the non-trivial inner product in (\ref{eq:overlap}) perturbatively in $1/N$ in terms of a classical interacting dimer model. 
To first nontrivial order, this model is exactly the one studied in \,\cite{Alet_dimers1,Alet_dimers2}, and so the Monte Carlo results in Fig.\ 26 of \cite{Alet_dimers2} can be utilized; in their notation $X_2=\alpha/2$ and $W=1+1/N$. This gives $\alpha_3 \simeq 1.4$, $\alpha_4 \simeq 1.52$, and $\alpha_5\simeq 1.6$, in addition to the already reported $\alpha_2\simeq 1.22$. Our numerical results therefore confirm that the leading-order term in this expansion gives accurate results. Note that higher order terms could slightly increase these values, as is also shown in the supplementary material of \cite{Damle}.

A presumably more-accurate value for $\alpha$ can be found relaxing the constraint $\ell \ll L$, and fitting the data to the full curve predicted by conformal field theory, not just the straight-line segment. While the two-point function on a torus is not fixed uniquely by conformal invariance, the appropriate form is known for the free-boson field theory for any $\kappa$ \cite{BigYellowBook}. Thus if we assume that the free-boson/Coulomb gas results still apply with a different $\kappa=1/\alpha$,  the $SU(2)$ dimer-dimer correlators are modified to 
\begin{equation}\label{eq:cft_fit}
 D_{||}(\ell_x)\sim f(\ell_x/L_x,\tau)^{-\alpha},
\end{equation}
with $\ell_x \gg 1$. $f$ is a universal function of the two dimensionless ratios $\ell_x/L_x$ and $\tau =i L_y/L_x$ and can be expressed in terms of a Jacobi theta function defined in \ref{sec:CFT_Jacobi}. We have 
\begin{equation}
 f(\ell_x/L_x,\tau)=\sum_{n=0}^{\infty}(-1)^n \sin \left[(2n+1)\pi \ell_x/L_x\right]e^{-\pi n(n+1)(L_y/L_x)}
\end{equation}
Notice that while this function reduces to $f(\ell_x/L_x,\tau)=\sin(\pi \ell_x/L_x)$ in the limit of a thin torus ($L_y/L_x \gg 1$), the approximation remains excellent for $L_y/L_x=1$. Performing the fits to Eq.~(\ref{eq:cft_fit}) in the same region $1\ll \ell_x \ll L_x$ as before (red solid curve in Fig.~\ref{fig:corr_su2_dimers}), we observe that the data reproduces very well the free-boson CFT result, even the upturn when $\ell_x$ is of order $L_x$. The fact that the CFT result seems to still hold here provides additional evidence in favor of universality. However, the exponents determined this way are slightly different; for example we find $\alpha_2\simeq 1.26$ in the odd case, and $\alpha_2\simeq 1.18$ in the even case. This small discrepancy with the previous $SU(2)$ results \cite{RVB1,RVB2}, also suggested by the large $N$ calculations\cite{Damle}, could possibly be resolved by studying larger systems.  

\begin{figure}[ht]
 \begin{center}
  \includegraphics[scale=0.8]{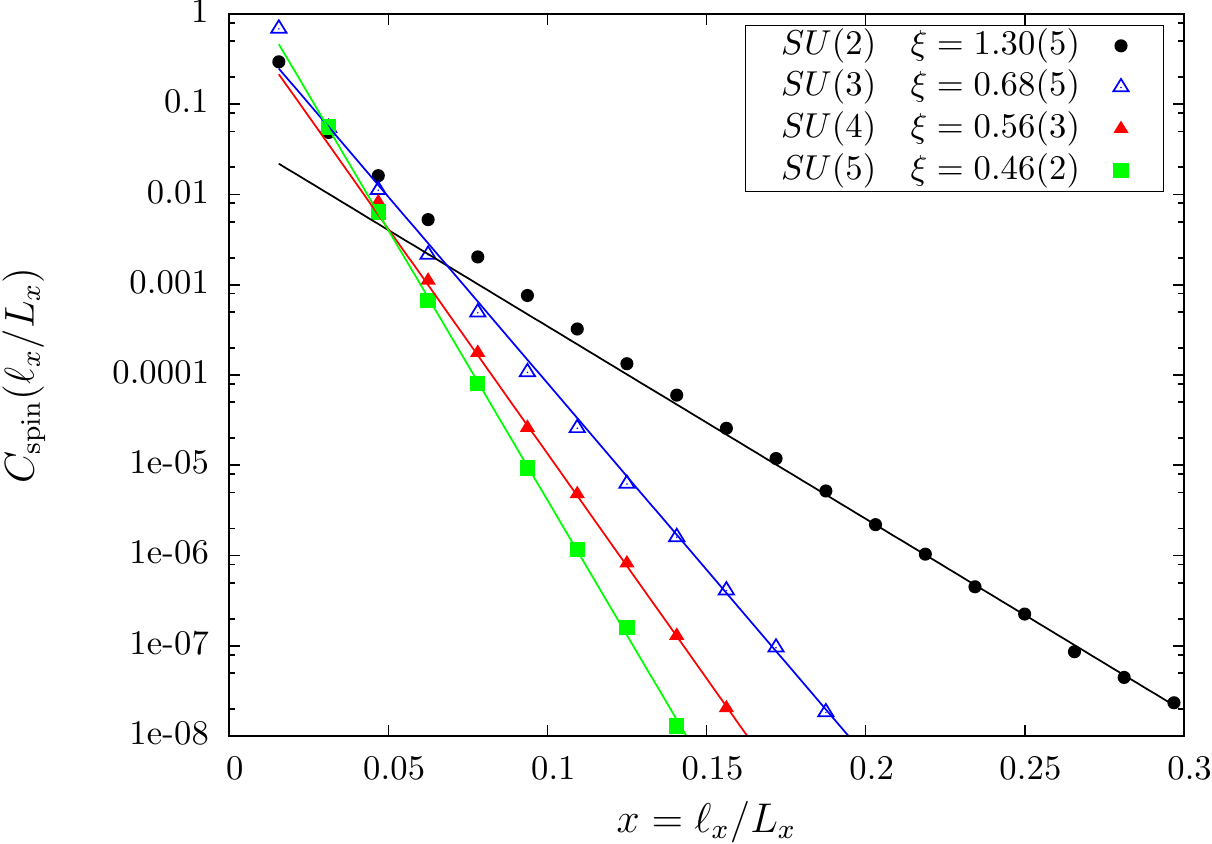}
 \end{center}
\caption{Spin-spin correlation function on a $64\times 64$ torus. We look at spins separated by a distance $\ell_x$ along the $x-$ direction:  $\mathbf{r}_1-\mathbf{r}_2=\ell_x \mathbf{e}_x$. }
\label{fig:corrspin_su2}
\end{figure}
We also studied spin correlations, which as discussed in section \ref{sec:RVB} are known to decay exponentially in the $SU(2)$ case: 
$
 \langle S(0)S(\ell_x) \rangle \sim \exp \left(-{\ell_x}/{\xi}\right)
$ \cite{LDA,RVB1}.
We find the same behavior in the $SU(N)$ case, with our results displayed in Fig.~\ref{fig:corrspin_su2}. We also observe that $\xi$ decreases with $N$, compatible with the intuition that the RVB state becomes more and more spin-disordered as $N$ increases, approaching zero correlation length in the dimer limit $N\to\infty$. Fitting the data gives  $\xi(N=2)=1.30(5)$ (compatible with \cite{RVB1}) and $\xi(N=3)=0.68(5)$, $\xi(N=4)=0.56(3)$ and $\xi(N=5)=0.46(2)$. This lack of long-range N\'eel order is consistent with quantum spin-liquid  behavior.

\section{Entanglement in the $SU(N)$ RVB wave function}
\label{sec:rvb_entanglement}

Throughout this paper we have emphasized the similarities between the $SU(N)$ and dimer RVB states. These make it plausible that the two-cylinder R\'enyi entropy for the $SU(2)$ RVB state in the scaling limit can be obtained from the free-boson result (\ref{eq:shape_general}). If this is true, this is a much stronger statement than the results for the critical exponent $\alpha$ as a function of $N$ in the previous section. The reason is that even though correlators in the classical dimer model can be obtained by taking the $N\to\infty$ limit of the loop model defined by the $SU(N)$ RVB inner product, the Hilbert space of the two is completely different; the latter being much larger. Thus a dimer crossing the boundary would carry much more entropy. Moreover, the reduced density matrix for $SU(N)$ has blocks with non-zero total spin, while these are by construction forbidden in the quantum dimer model. Thus it is very unlikely that non-universal quantities will be the same in the two cases.

All our results are consistent with the subleading part of the two-cylinder
R\'enyi entropy being universal. Thus it is possible that in the scaling limit,
this piece still could be given by the conformal field theory result, with the
identification of $\kappa$ coming from the dimer correlation functions results.
In this section we provide evidence for such universality by finding the curves
for arbitrary $\alpha$, and extending previous numerical results \cite{Ju2012}
for $SU(2)$ to the $SU(N)$ case.

\begin{figure}[t]
 \begin{center}
  \includegraphics[scale=0.4]{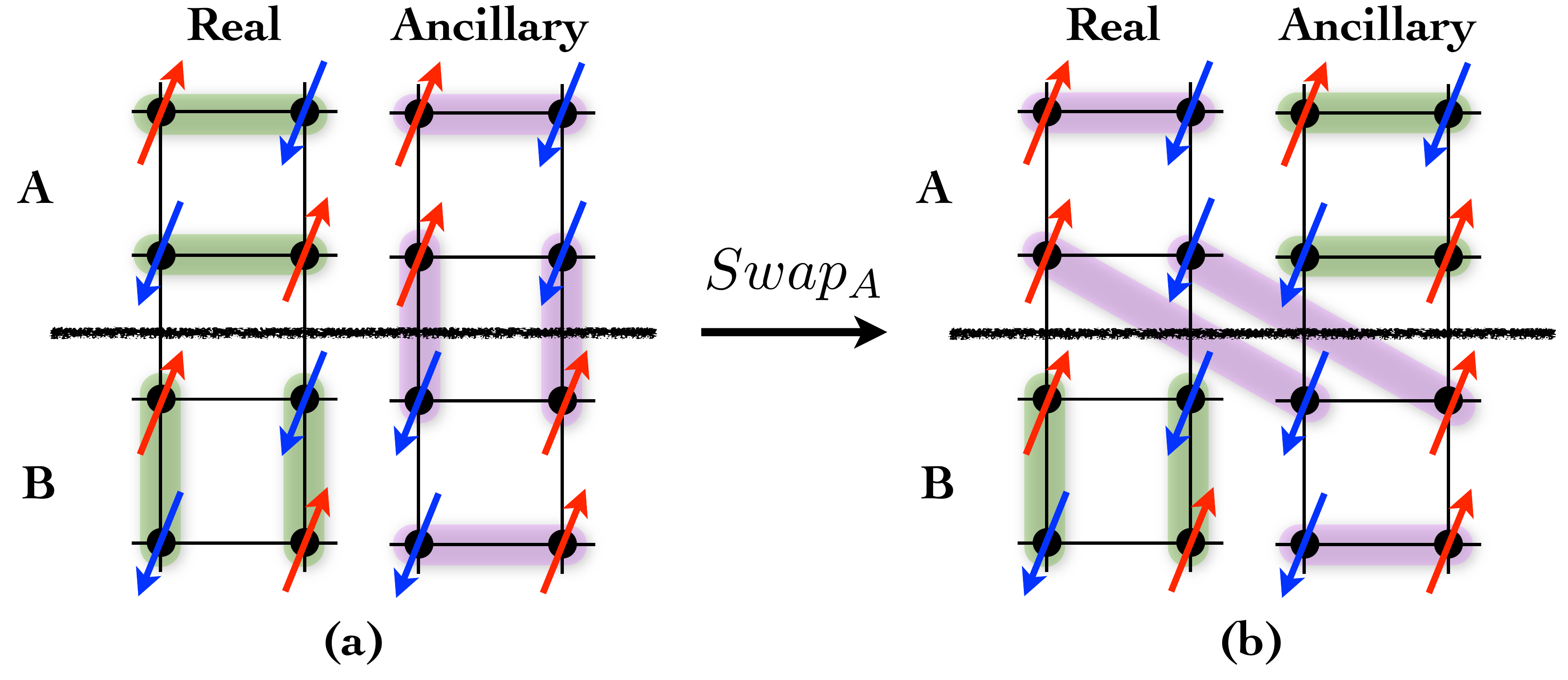}
 \end{center}
\caption{The $Swap_A$ operator. In (a), we show the original state, $|V_{\mathcal C} \rangle$, which is composed of a physical (or ``real'') system of eight lattice sites, and a non-interacting replica (or ``ancillary'').
In (b), we show the result of applying the $Swap_A$ operator, $Swap_A |V_{\mathcal C} \rangle =  |V_{\mathcal C^{\star}} \rangle$. When acting the operator on this state, it is important to note that {\it both} the spins and bond bases in region A get swapped between the real and ancillary copies.}
\label{fig:swap}
\end{figure}

\subsection{Monte Carlo evaluation of the R\'enyi entropies}
\label{sec:su2_numerics}

In order to calculate the R\'enyi entropies, the Monte Carlo sampling algorithm of Section \ref{sec:numerics} must be modified to incorporate
a {\it replica trick} procedure, first described in Ref.~\cite{swap}.  Namely, in order to calculate the $n$-th order R\'enyi entropy Eq.~\ref{RenyiEq}, one measures the expectation value of a $Swap_A$ (or permutation) operator acting between $n$ copies of the system.  For the $SU(N)$ wavefunctions studied here, we restrict ourselves to $n=2$.  

A naive sampling of the $Swap_A$ operator, shown graphically in Fig.~\ref{fig:swap}, is straightforward.  One simply identifies the spatial region $A$ and its complement $B$, then applies the operator to a sampled RVB state, $Swap_A |V_{\mathcal C} \rangle =  |V_{\mathcal C^{\star}} \rangle$ (where each state $|V_{\mathcal C} \rangle$ contains $n=2$ replicas).  The second R\'enyi entropy can be calculated from \cite{swap}
\begin{equation}
	\rho_A^2
		= \frac{\langle V_{\mathcal{L}} | V_{\mathcal{R}^{\star}} \rangle}
			{\langle V_{\mathcal{L}} | V_{\mathcal{R}} \rangle} 
			 = N^{n_{\rm swap} - n_l}\ ,
\end{equation}
where $n_l$ is the number of loops in the transition graph of the two un-swapped RVB states (the denominator), $n_{\rm swap}$ is the number of loops in the transition graph using one swapped state, and $N$ denotes the number of spin flavors from $SU(N)$.  
Thus, the reconfiguration of the valence-bond endpoints is the important ingredient of the $Swap_A$ operator for this measurement.  Note, 
since the $Swap_A$ operator serves to swap {\it all} basis states in the spatial region $A$, spin states $|Z_{\mathcal C} \rangle$ can effectively be ignored when performing this measurement; they are only used as a condition to {\it sample} the original un-swapped VB configurations.

However, one discovers quickly as the lattice size (and particularly the size of region $A$) grows, sampling statistics become exponentially poor when using the naive $Swap_A$ operator to calculate $S_2$.  One must therefore employ some variant of the {\it ratio} trick described in  Ref.~\cite{swap}.  As discussed in detail there, this involves sampling 
\begin{equation}
 \frac{\langle V_{\mathcal{L}} | Swap_A | V_{\mathcal{R}} \rangle}
			{\langle V_{\mathcal{L}} | Swap_X | V_{\mathcal{R}} \rangle},
\end{equation}
where $X$ is a spatial region smaller than $A$, i.e.~$X \subsetneq A$.  In order to implement this measurement in the RVB wavefunction, a completely different simulation must be run with the overlap ${\langle V_{\mathcal{L}} | Swap_X | V_{\mathcal{R}} \rangle}$ occurring in the weight.  

The essential point of the ratio trick then is that simulations must be run with the basis states corresponding to region $X$ {\it already} swapped before measurement of the $Swap_A$ operator.  Since we work in a combined VB-spin basis, one imagines the simulation beginning with 
two states, $|V_{\mathcal L} \rangle = |D_{\mathcal L} \rangle | Z_{\mathcal L} \rangle$ and $|V_{\mathcal R'} \rangle = |D_{\mathcal R'}  \rangle | Z_{\mathcal R'} \rangle$.  However, the simulation is performed with ${\langle V_{\mathcal{L}} | Swap_X | V_{\mathcal{R}} \rangle}$ in the weight, 
meaning that $ |D_{\mathcal R'} \rangle | Z_{\mathcal R'} \rangle = Swap_X |D_{\mathcal R} \rangle | Z_{\mathcal R} \rangle  $.  Therefore, it is important to sample spin states such that the overlap $\langle Z_{\mathcal L} | Z_{\mathcal R '} \rangle$ is nonzero.  

Note that, when performing the Monte Carlo update on the valence-bond state $| D_{\mathcal{R} '} \rangle$, one does not have to consider the overlap of $\langle V_{\mathcal{L}} | V_{\mathcal{R}'} \rangle$ since the overcompleteness condition ensures that the inner product of new valence-bond configurations (with the same spin configuration) are always non-zero.  Hence, it may be convenient to consider the ``un-swapped'' state $|V_{\mathcal R} \rangle$ for the purposes of sampling the VB configuration, where one can straightforwardly perform local (plaquette-flip) or non-local (loop) updates on the VB state as described in 
Section \ref{sec:numerics}.

We employ ratio-trick evaluations of the $Swap_A$ operator to calculate the scaling of the second R\'enyi entropy in the $SU(N)$ RVB wavefunction.  In the next subsection, we compare our results from these Monte Carlo simulations to those obtained from conformal field theory.

\subsection{Comparison of numerical results and CFT}

Here we compare our quantum Monte Carlo results with curves found by using Eq.~(\ref{eq:shape_general}) with partition functions coming from conformal field theory for arbitrary stiffness $\kappa=1/\alpha$. This two-cylinder entropy is that of the quantum Lifshitz state, the continuum limit of the quantum six-vertex state. If the universality discussed above holds true, they will also apply to the continuum limit of the square-lattice $SU(N)$ RVB states, with appropriate identification of $\alpha$ in terms of $N$.

\begin{figure}[ht]
 \begin{center}
  \includegraphics[width=8cm]{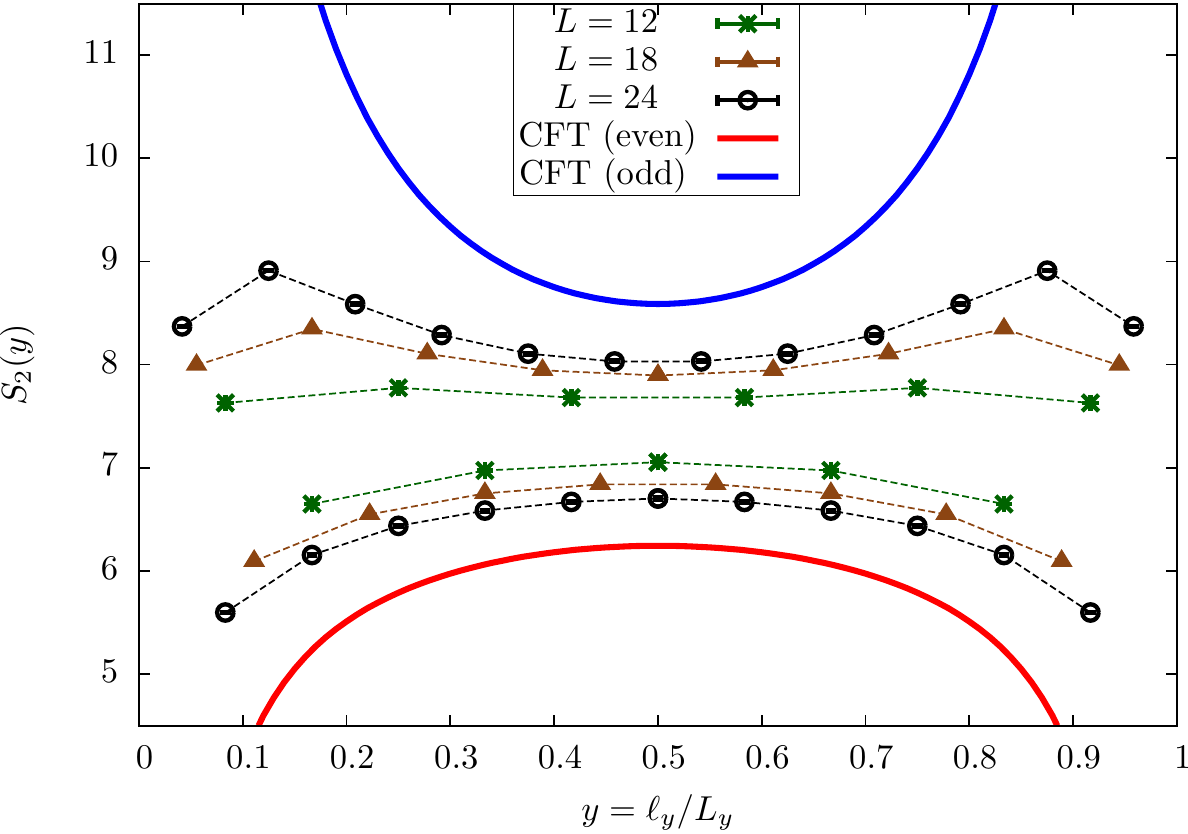}
  \includegraphics[width=8cm]{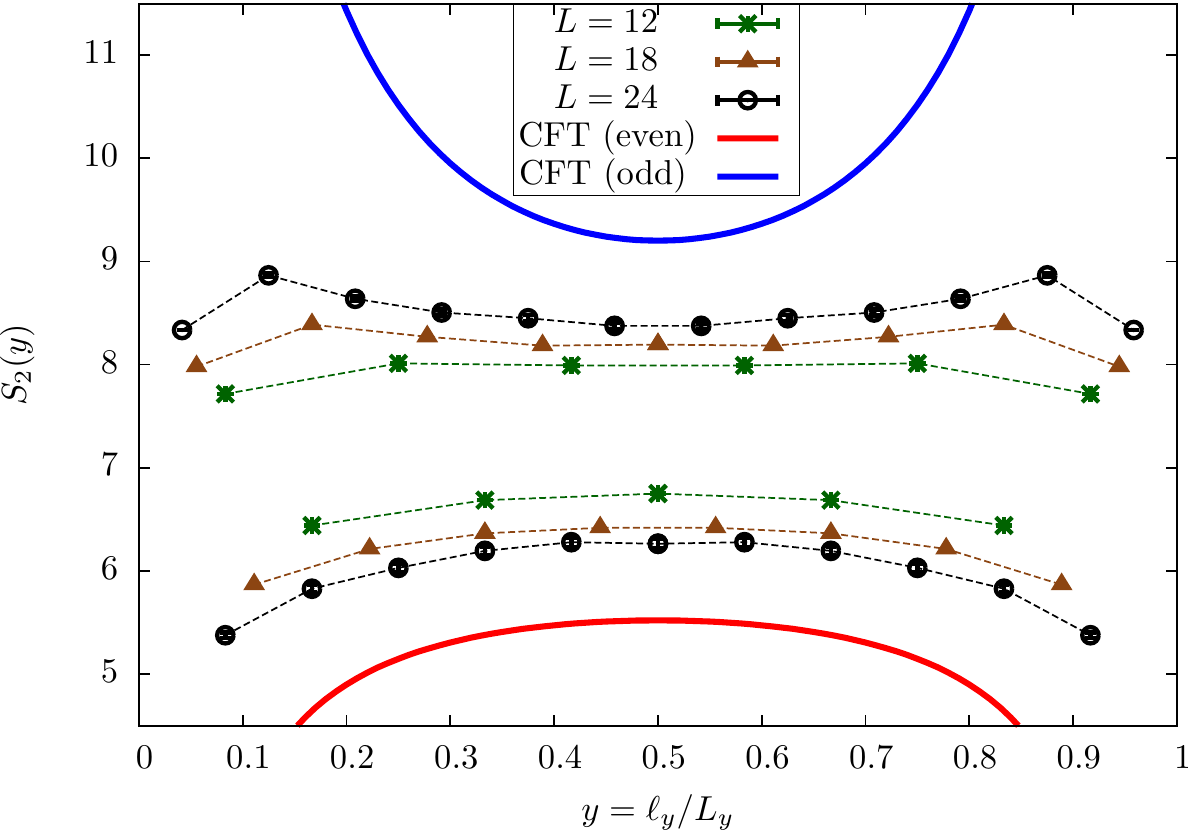}
 \end{center}
\caption{Numerical extraction of the universal shape of $S_2$ as a function of $y=\ell_y/L_y$ for the $SU(2)$ RVB wave function and $L=L_x=L_y=12,18,24$. Red and blue curves are the CFT results, using Eqs.~(\ref{eq:cft_prediction_gen},\ref{eq:cft_prediction_gen_odd},\ref{eq:cft_w0_gen},\ref{eq:cft_w0_gen2}). \emph{Left:} $W=all$ sector.  \emph{Right:} $W=(0,0)$ sector. We observe in this case a stronger even-odd effect on the lattice, consistent with the expected result in the scaling limit.}
\label{fig:SU2_shape}
\end{figure}

Let us begin with the $SU(2)$ case, and present our results for $S_2$ both for unrestricted windings and for the $W=(0,0)$ winding sector. The quantum Monte Carlo data is shown in Fig.~\ref{fig:SU2_shape}.
We observe a strong similarity to the dimer data. In particular, there is a strong even-odd effect that becomes even bigger in the $W=(0,0)$ sector.
This suggests $S_2$ for the $SU(2)$ RVB state lies in a locked phase, just as its dimer counterpart does. 
This scenario is also supported by exact diagonalizations on small systems, which show that the biggest eigenvalue of the reduced density matrix is non-degenerate:  $\Delta_{SU(2)}\equiv -\ln (p_1/p_{\rm max}) \approx 2.67$ for a $4\times 4$ torus cut in two halves, and therefore $d_{\rm min}=1$. (The exact value for dimers is $\Delta_{\rm dimers}=2\ln \pi\simeq 2.29$ in the scaling limit \cite{Stephan2012}.) Such a large entanglement gap has also been observed recently for the $SU(2)$ RVB state, in the closely related infinite-cylinder geometry \cite{Poilblanc}. We also refer to \cite{Poilblanc} for a detailed study of the entanglement spectrum in this geometry.

With no restrictions on the winding numbers, the conformal field theory result for the two-cylinder R\'enyi entropy at general $\alpha$ is
\begin{eqnarray}\label{eq:cft_prediction_gen}
 s_n^{\rm (even)}(y,\tau)&=&\frac{n}{1-n}\ln \left(\frac{\alpha}{2}\times\frac{\eta(\tau)^2}{\theta_3(\alpha\tau)\theta_3(\tau/\alpha)}\times\frac{\theta_3(\alpha y \tau)\theta_3(\alpha(1-y)\tau)}{\eta(2y\tau)\eta(2(1-y)\tau)}\right)\ ,\\ 
 \label{eq:cft_prediction_gen_odd}
 s_n^{\rm (odd)}(y,\tau)&=&\frac{n}{1-n}\ln \left(\frac{\alpha}{2}\times\frac{\eta(\tau)^2}{\theta_3(\alpha\tau)\theta_3(\tau/\alpha)}\times\frac{\theta_4(2y \tau)\theta_4(2(1-y)\tau)}{\eta(2y\tau)\eta(2(1-y)\tau)}\right)\ .
\end{eqnarray}
The prediction of Eq.~(\ref{eq:cft_w0}) for the $W=(0,0)$ winding sector is generalized to:
\begin{eqnarray}\label{eq:cft_w0_gen}
 s_{n,0}^{\rm (even)}&=&\frac{n}{1-n}\ln \left(\frac{\alpha}{2}\frac{\eta(\tau)^2\theta_3(\alpha y(1-y)\tau)}{\eta(2y\tau)\eta(2(1-y)\tau)}\right)\ , \\
 \label{eq:cft_w0_gen2}
 s_{n,0}^{\rm (odd )}&=&\frac{n}{1-n}\ln \left(\frac{\alpha}{2}\frac{\eta(\tau)^2\theta_4(\alpha y(1-y)\tau)}{\eta(2y\tau)\eta(2(1-y)\tau)}\right)\ .
\end{eqnarray}
We have made the assumption that the boundary conditions are the same as those for dimers (i.e.\ $a=0,1/2$ for even and odd respectively for all $\alpha$).

The curves (\ref{eq:cft_prediction_gen}) and (\ref{eq:cft_prediction_gen_odd}) 
 with $\alpha=1.2$ are plotted in fig.~\ref{fig:SU2_shape} along with the $SU(2)$ Monte Carlo data.  The trend in the data is clearly toward these curves, but as with the dimer case, the finite-size effects are very strong. 
Several additional interesting features become apparent by plotting the CFT curves
(\ref{eq:cft_prediction_gen}) and (\ref{eq:cft_prediction_gen_odd}) at $\alpha=2$ and $\alpha=1.2$, corresponding to the dimer and $SU(2)$ RVB exponents respectively.
\begin{figure}[ht]
 \begin{center}
  \includegraphics[width=8cm]{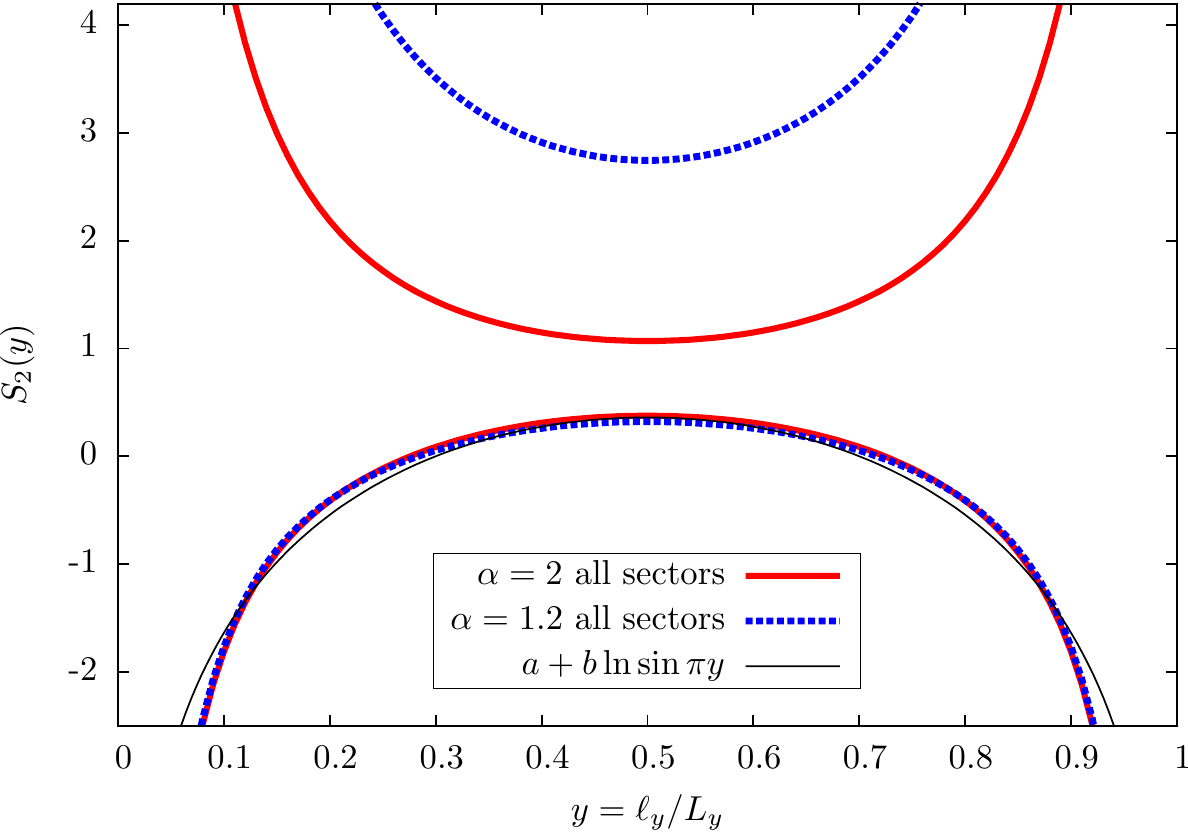}
    \includegraphics[width=8cm]{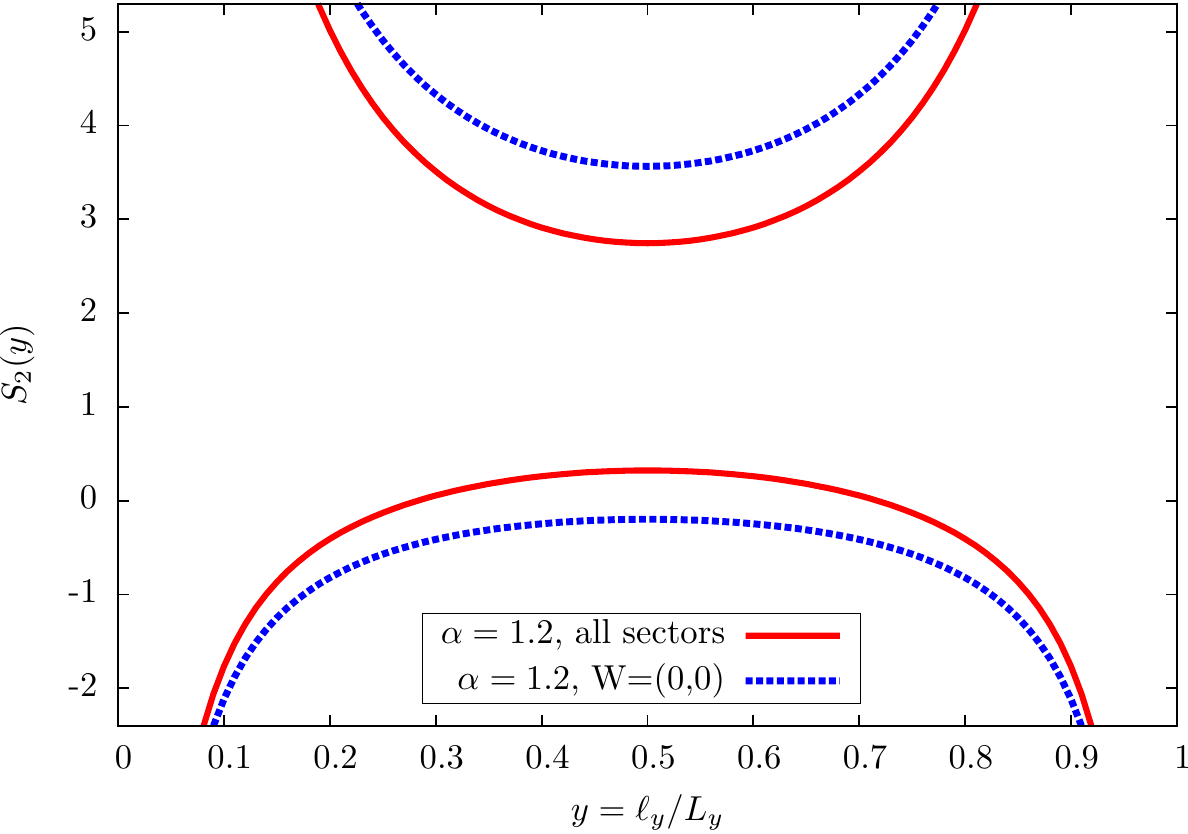}
 \end{center}
\caption{CFT predictions for the universal shape of $S_2$, as a function of $y=\ell_y/L_y$.\emph{Left:} Red solid lines correspond to the dimer RVB state with $\alpha=2$, while the blue dashed lines are the $SU(2)$ RVB case, with $\alpha= 1.2$. We also plot in black the usual $1d$ result $a+b\ln \sin \pi y$ with appropriate $a$ and $b$. \emph{Right: }Prediction for $\alpha=1.2$. Blue dashed lines is the $W=(0,0)$ sector, while red solid lines show the prediction with no restriction on winding numbers.}
\label{fig:SUN_CFT}
\end{figure}
As is apparent from fig.~\ref{fig:SUN_CFT}(left), the curve for the even branch is essentially independent of the $\alpha$ over a large range, and resembles the $1d$ result $S_2\propto\ln \sin (\pi y)$. The shape of the odd curve only weakly depends on $\alpha$, but the curve does appreciably shift in this range. We also observe --fig.~\ref{fig:SUN_CFT}(right)-- that contrary to the dimer case ($\alpha=2$), the two odd curves corresponding to $\alpha=1.2$ are not nearly identical in the $W=(0,0)$ and $W=all$ sectors. This feature is also in qualitative agreement with the data shown in fig.~\ref{fig:SU2_shape}.

Also plotted in fig.~\ref{fig:SUN_CFT} is the curve $a+b\ln (\sin \pi y)$, with $a$ and $b$ fitted to make the curves as close as possible. This curve is the 1d R\'enyi entropy for a conformal field theory with space a circle of circumference $L$ split into segments of length $yL$ and $(1-y)L$. The logarithmic term  is the leading term in one dimension, with its coefficient $b$ proportional to the central charge \cite{Cardy}. 
One might expect that our 2d curve would resemble the 1d curve when the 2d system is obtained by coupling together critical 1d systems while maintaining the criticality. This for example is the case for fermions with $\pi$ flux discussed in \cite{Ju2012}. However, the square-lattice quantum dimer model is gapless only in the two-dimensional limit (just like the Goldstone modes for the Heisenberg model also discussed in \cite{Ju2012}), so we see no particular reason for the curves to be so similar. As can be seen from the figure, the two curves are quite close,  especially if one considers 
the numerical accuracy with which they can usually be determined. They are, however, clearly not identical.

\begin{figure}[ht]
 \begin{center}
  \includegraphics[width=10cm]{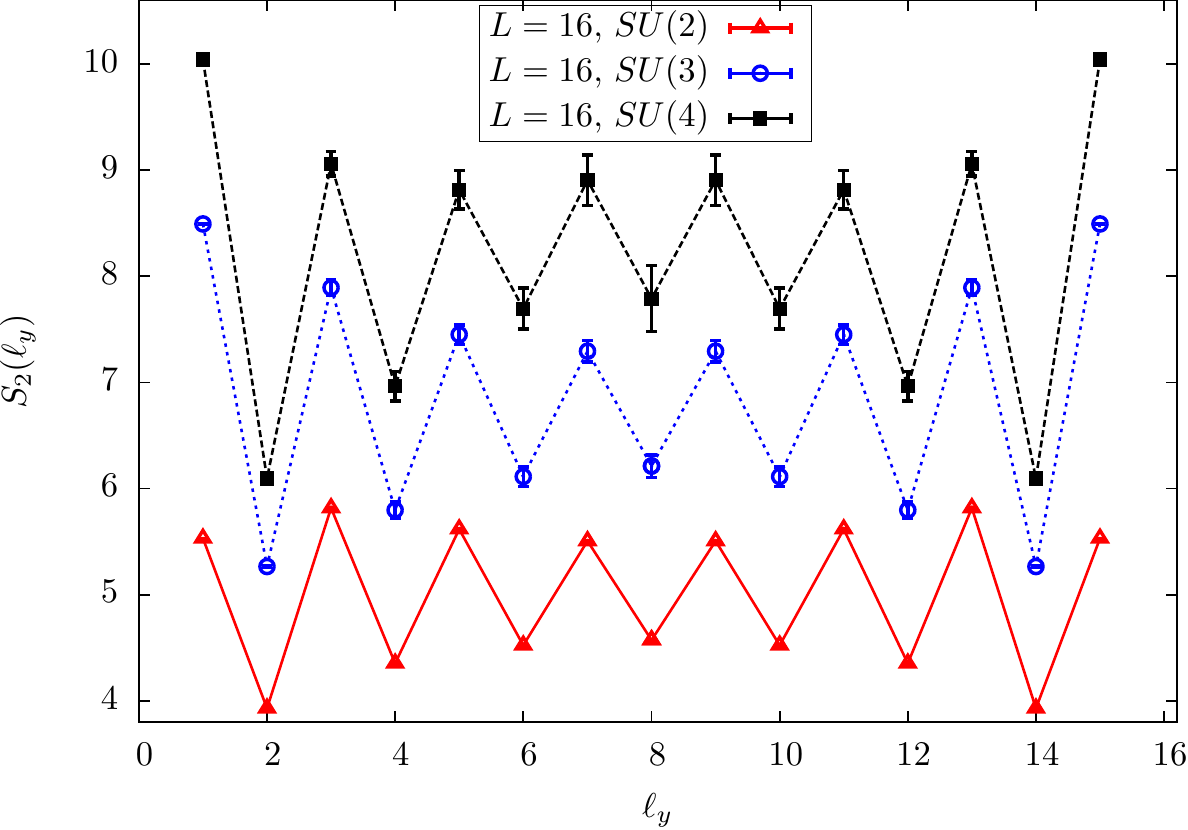}
 \end{center}
\caption{Numerical data for the second R\'enyi entropy $S_2$ in the $SU(2,3,4)$ case. As can be seen, even on relatively small system sizes the numerical data is not totally converged. }
\label{fig:SUN_shape}
\end{figure}

The same picture still seems to hold for the $SU(N)$ RVB state. We plot $S_2$ data for $L=16$ and $SU(2,3,4)$ is shown in Fig.~\ref{fig:SUN_shape}. As can be seen, the error bars get bigger and bigger for larger $N$, but we still observe a clear even-odd effect. This strongly suggests that we are still in a locked phase, and that Eq.~(\ref{eq:cft_prediction_gen}) applies in the thermodynamic limit. 

Extracting an accurate exponent $\alpha_N$ from the $S_2$ data is difficult, due to the combined finite-size effects and statistical errors. A possible try would be to  look at the even-odd difference for $y=1/2$. On the lattice it is defined as $\delta_n(L_x,L_y)=\left|S_n(L_y/2+1)-S_n(L_y/2)\right|$ for even $L_y$. This difference should go to a positive constant in the thermodynamic limit. From Eq.~(\ref{eq:cft_prediction_gen}) we predict
\begin{equation}\label{eq:even_odd}
 \delta_n(\tau,\alpha)=\frac{2n}{1-n}\ln \left(\frac{\theta_4(\alpha\tau/2)}{\theta_3(\alpha \tau/2)}\right)
\end{equation}
Using our biggest system-size even-odd data in (\ref{eq:even_odd}) yields $\alpha \approx 2.27$ for dimers and $\alpha \approx 1.57$ for $SU(2)$ RVB. These numbers can be improved by extrapolation, but the strong finite-size effects described above prevent an accurate estimate of these numbers.

With such strong finite-size effects, we cannot make a definitive conclusion. Our results, however, provide qualitative evidence for  Eq.~(\ref{eq:shape_general}) applying to the $SU(2)$ RVB state. They also support the convention of \cite{Ju2012} that the subleading piece of the two-cylinder R\'enyi entropy is universal.

Unfortunately, we do not know how to derive these results as we did for the dimers. The mapping of Ref.~\cite{Shannonee} to a classical R\'enyi-Shannon entropy is not valid for the $SU(N)$ RVB states; because of the non-orthogonality of different valence-bond configurations, the Schmidt decomposition of the RVB state is much more complicated.  Moreover, a naive extension of the result of \cite{Stephan2011} would then yield a critical value $ n_c=\alpha/2\simeq 0.6$  for $SU(2)$. This would imply an even-odd effect even for the von Neumann entropy, in violation of strong subadditivity \cite{Strongsubadditivity}. The formula $n_c=d_{\rm min }^2/(2\kappa)=d_{\rm min}^2\alpha/2$ therefore cannot hold in general for models that cannot be mapped onto a classical R\'enyi-Shannon entropy. The fact that it does apply to quantum six-vertex states does \emph{not} 
contradict strong subadditivity, because in this case $d_{\rm min}=2$ and $\alpha \geq 1$, so $n_c\geq 1$ in this case. \footnote{Notice that naively  applying $n_c=d_{\rm min}^2 \alpha/2$ to an RK wave function built out of interacting dimers \cite{Alet_dimers1,Alet_dimers2} would, as with the $SU(N)$ RVB, violate strong subadditivity.} 

A possible way to test our CFT results while sidestepping this difficulty would be to look at another quantity, defined as follows. Consider another $SU(2)$ RVB wave function $|\Psi^\prime\rangle$ living on the torus, but where all valence bond states $|V_{\cal C}\rangle$ with singlets crossing the boundaries between $A$ and $B$ have been removed. Such a wave function can be rewritten as
the tensor product of two RVB-wave functions living respectively on cylinders $A$ and $B$:
\begin{equation}
 |\Psi^\prime\rangle =|\Psi(A)\rangle \otimes |\Psi(B)\rangle\ .
\end{equation}
 Consider then the logarithmic (bipartite) fidelity \cite{Bipartite_fidelity} defined by taking the scalar product with the original (torus) RVB wave function:
\begin{equation}\label{eq:lbf}
 F=-\ln \left(\left|\langle \Psi|\Psi^\prime\rangle\right|^2\right).
\end{equation}
This fidelity enjoys several properties very similar to the entanglement entropy, like a generic area law, and a universal logarithmic divergence for one-dimensional quantum critical points. Using Eq.~(\ref{eq:nonortho}), it can be expressed using classical partition functions for the RVB loop gas. In this case loops longer than the dimers can cross the boundary between $A$ and $B$, but since the dimers dominate, we expect our main result (\ref{eq:shape_general}) to apply to this quantity, upon removing the $n/(n-1)$ factor (or equivalently setting $n=\infty$). Using the arguments of section \ref{sec:bpt}, one can also check that the equality
\begin{equation}\label{eq:renyilbf}
F=S_\infty 
\end{equation}
holds exactly on the lattice for the dimer RVB state. While this is not true anymore for $SU(N)$ RVB states, we expect (\ref{eq:renyilbf}) to remain true at the level of universality. Therefore, Eqs.~(\ref{eq:cft_prediction_gen},\ref{eq:cft_prediction_gen_odd}) should describe the shape of both $S_\infty$ and $F$, when varying the respective sizes of the cylinders $A$ and $B$. Studying this fidelity, for example using quantum Monte Carlo methods similar to the ones presented here, could probably shed some more light on the phase transition scenario.

\section{Conclusion} 
\label{sec:conclusion}


In this paper we studied a family of resonating-valence-bond (RVB) states on the square lattice, focusing in particular on the R\'enyi entropy $S_n$ for dividing a torus into two cylinders. This provides an elegant probe into the physics of such gapless states. We found several interesting characteristics of their quantum entanglement, by deriving exact results for the dimer and quantum Lifshitz RVB states, and by numerical work on the $SU(2)$ and the $SU(N)$ RVB states on the square lattice. 

One of our main results is confirmation that the R\'enyi entropy for RVB states exhibits a two-branch structure, when the length of the entangled region varies between an even and odd number of lattice sites. We computed this exactly for the dimer RVB functions by mapping the R\'enyi entanglement entropy to a classically-computable R\'enyi-Shannon entropy, and demonstrated that the two-branch structure exists for {$n>1$ (but not for $n\leq 1$)}.
By measuring the Monte Carlo expectation value of a ``swap'' operator on a replicated lattice, we find $S_2$ for the $SU(N)$ RVB state for $N=2,3,4$, and find the same two-branch structure. It would be interesting to extend the Monte Carlo calculations to higher $n$, where the replica trick can still be used, {and to study the fidelity of Ref.~\cite{Bipartite_fidelity} in this context.} 

The two-branch shape of the R\'enyi entropies has an interesting consequence:  strong subadditivity is violated for the ``odd'' branch, which is a concave-up function of the linear size of the entangled region.  We demonstrate that this is far from a numerical artefact even away from the dimer case, by calculating the shape of these branches in an effective conformal field theory.  This effective theory involves partition functions that can be evaluated exactly in the continuum, and exhibit the same two-branch structure observed in lattice calculations.  This emphasizes the universality of the result and confirms that the two-branch R\'enyi entropies are not the result of finite-size lattice effects, {contrary to what generically happens in one-dimensional systems \cite{Corrections_scaling1,Corrections_scaling2}.} 

From this analysis, we have elucidated two important points regarding the shape of the R\'enyi entropy as a function of linear size of the entangled region.  
First, the two-branch structure does not occur for every R\'enyi index $n$, but only for values {strictly} larger than some (non-universal) $n_c$, which must be greater than unity since the von Neumann entropy $S_1$ is strictly constrained by strong-subadditivity (while the $n>1$ R\'enyi entropies are not).  We believe that in the square-lattice RVB, {$1\leq n_c<2$}; since our Monte Carlo methods require $n\ge 2$, they will always find a two-branch structure as observed here.  
It would be useful to change $d_{\rm min}$ by studying the RVB wave function on other bipartite lattices. For example on the honeycomb lattice $d_{\rm min}=3$ (and therefore $n_c=3^2=9$) for dimers, and $S_2$ for RVB would likely lie in the replica phase, {and would not give rise to an even-odd effect.} 
This prediction should be tested by future Monte Carlo studies.

Second, we have found an exact expression for the shape of the R\'enyi entropy curves in a geometry where a linear size of the entangled regions grows as {$\ell$}.  The fact that it is consistent with our numerical data for the $SU(2)$ and dimer RVB wavefunctions is a strong indication that  it is universal. We see no particular reason why this universality should be a characteristic only of RVB wave functions, so this conjecture should be tested on future gapless wavefunctions.  Indeed, DMRG results for the 2d transverse-field Ising model critical point have already exhibited the same type of behavior \cite{Konik}.
As observed in previous studies of gapless systems \cite{Ju2012}, these curves are well approximated by the celebrated universal one-dimensional result {$\ln(\sin(\pi \ell /L))$}; however, deviations were observed.  {Our exact 2d results from effective field theory indeed confirm that the 1d result is only approximate in the RVB case}, but also that it is a very good approximation.

We have also explored the crossover between the $SU(2)$ and dimer RVB states by studying the $SU(N)$ states. For equal-time correlators, it is possible to interpolate between the $SU(2)$ and dimer states by taking $N \rightarrow \infty$, but there is no guarantee that there is no transition at some value of $N$.  Previous studies demonstrated that the $SU(2)$ RVB state exhibited algebraic decay with an exponent $\alpha_2 \approx 1.2$, while the dimer RVB has $\alpha_{\infty} = 2$.  We showed in this work that this exponent evolves smoothly as $N$ is varied, with no evidence for a transition.  
This is strong evidence of universality, allowing us to link various physical aspects of the $SU(2)$ case with exact results on the dimer RVB. Perhaps even more importantly, we also showed that this correspondence at least qualitatively holds for the putatively universal part of the two-cylinder R\'enyi entropy as well. This to us is evidence both for the universality and the effectiveness of using dimer models to probe at least some {\it highly non-trivial} aspects of the $SU(2)$ RVB state.

It is worth recalling that the $SU(N)$ and dimer RVB states are the simplest wavefunctions exhibiting features of gapless spin-liquid behavior. Since gapless spin liquids are long-range entangled states {\it not} amenable to  classification by simple quantities such as the topological entanglement entropy, we hope that some of the universal scaling properties uncovered here will be of use in future efforts to classify and characterize them in theory, numerics, and experiment.

\paragraph{Acknowledgments}
 We would like to thank Matthew Hastings for early collaboration on this project. We also thank  A.\ Kallin, H. Casini, E.\ Fradkin, G.\ Misguich, V.\  Pasquier, A. Sandvik, R. Myers, and X-G. Wen for stimulating discussions. This work has been supported by the Natural Sciences and Engineering
Research Council of Canada (NSERC), and by the US NSF via grants DMR/MPS-0704666 and DMR/MPS1006549.   Simulations were performed on the computing facilities of SHARCNET.

 \appendix
 \clearpage

\section[\;\;\;\;\;\;\;\;\;\;\;\;\;\;R\'enyi-Shannon entropy on the torus]{R\'enyi-Shannon entropy for dimers on the torus}
\label{sec:lgv}
We wish to compute the following classical entropy
\begin{equation}
 S_n=\frac{1}{1-n} \ln \left(\sum_{\sigma,\mu} [p_{\sigma,\mu}]^n\right), 
\end{equation}
for any real $n$. The probabilities $p_{\sigma,\mu}$ we are interested in are given by a ratio of partition functions, which can handily be expressed using a transfer matrix
\begin{equation}\label{eq:tm}
 p_{\sigma,\mu}=\frac{Z_{\sigma,\mu}^A Z_{\mu,\sigma}^B}{Z}=\frac{\langle  \sigma|T^{\,\ell_y}|\mu\rangle\langle \mu|T^{\,L_y-\ell_y}|\sigma\rangle}{{\rm Tr}\; T^{L_y}}
\end{equation}
$T$ is the transfer matrix of the dimer model, and acts on the vector space generated by dimer occupancies on vertical edges along a horizontal line: $\langle a|T|b\rangle=1$ if configuration $|a\rangle$ and $|b\rangle$ are compatible, $0$ otherwise. The denominator is calculated in \ref{sec:dimers_exact}. Each factor on the numerator of Eq.~(\ref{eq:tm}) can be evaluated using free fermions techniques\cite{Lieb1967,Alet_dimers2,Shannonee}. 

\begin{figure}[ht]
\begin{center}
\includegraphics{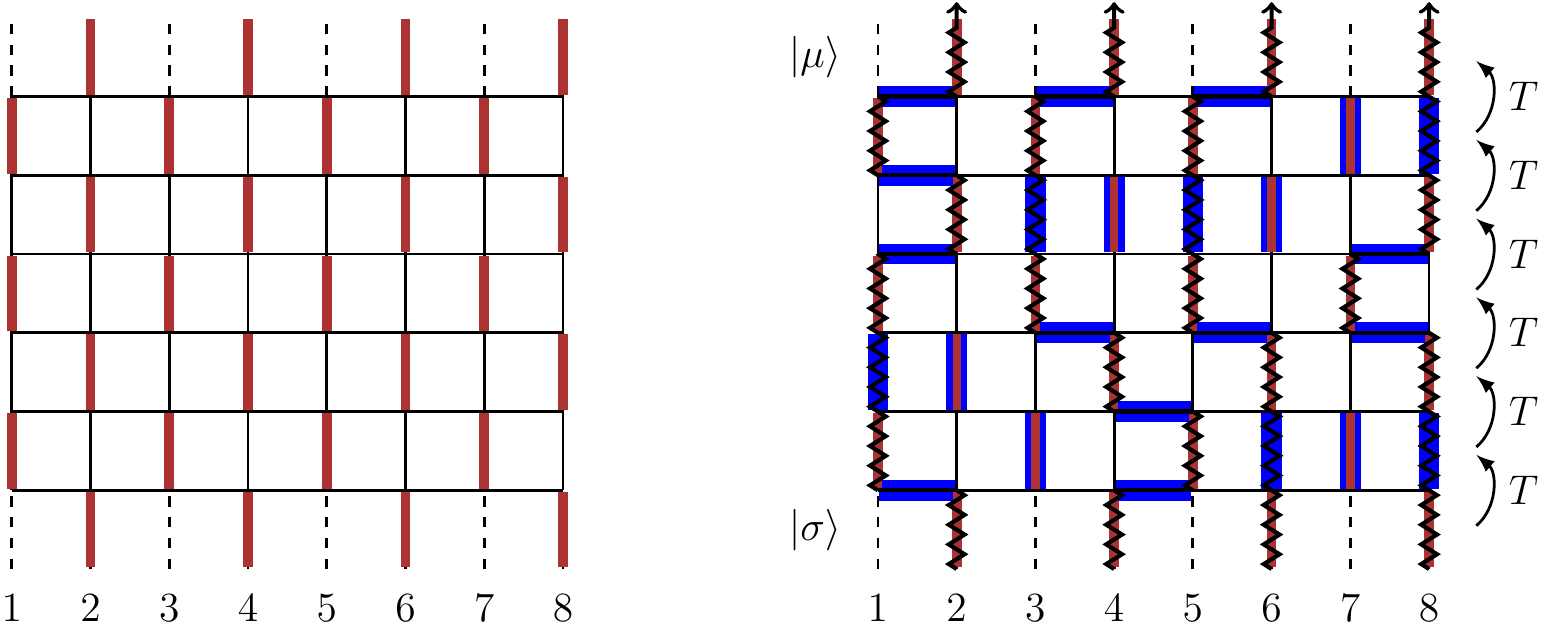}
  \end{center}
  \caption{Illustration of the mapping onto free fermions for open boundary conditions. \emph{Left:} reference configuration with staggered dimers. \emph{Right:}  transition graph generated by superimposing the reference configuration on a given dimer configuration (in thick blue). This generates a collection of $4$ non-intersecting lattice paths going upwards, which may be expressed in terms of fermions (black zigzag lines). Here $|\sigma\rangle=c_2^\dag c_4^\dag c_6^\dag c_8^\dag|0\rangle$.}
  \label{fig:freefermions}
  \end{figure}

The mapping goes as follows. Let us first consider a particular configuration with staggered dimers, which we call reference configuration (see fig.~\ref{fig:freefermions} on the left). Superimposing the reference on any other dimer configuration generates a collection of non-intersecting lattice paths, represented by black lines in fig.~\ref{fig:freefermions}(right). 

There is a one to one correspondence between the dimer configurations and the lattice paths, provided the latter obey a certain set of rules. To clarify them, it is most convenient to reformulate everything in terms of fermions, because their statistics will naturally encode the non-intersection constraint. 
Define a fermion as a vertical link occupied by a reference (red) dimer only, or a real (blue) dimer only. These are represented by black zigzag lines in fig.~\ref{fig:freefermions}. $|\sigma\rangle$ and $|\mu\rangle$ can be rewritten using fermion creation operators as
\begin{eqnarray}
|\sigma \rangle&=&|x_1,x_2,\ldots,x_N\rangle=c_{x_1}^\dag c_{x_2}^\dag \ldots c_{x_N}^\dag |0\rangle \\
|\mu \rangle&=&|y_1,y_2,\ldots,y_N\rangle=c_{y_1}^\dag c_{y_2}^\dag \ldots c_{y_N}^\dag |0\rangle
\end{eqnarray}
where the $x_i$ and $y_j$ label the ordered positions of the $N$ fermions. Introducing of shift of $+1$ lattice spacing from one row to the other, it is easy to check from fig.~\ref{fig:freefermions} that the transfer matrix satisfies
\begin{eqnarray}\label{eq:tm1}
T|0\rangle&=&|0\rangle\\\label{eq:tm2}
T c_{2j-1}^\dag T^{-1}&=&c_{2j}^\dag\\\label{eq:tm3}
T c_{2j}^\dag T^{-1}&=&c_{2j}^\dag+c_{2j+1}^\dag +c_{2j+2}^\dag
\end{eqnarray}
Eqs.~(\ref{eq:tm1},\ref{eq:tm2},\ref{eq:tm3}) conserve the number of fermions, and encode the dimer close-packed and hard-core constraints on the square lattice. To account for periodic boundary conditions  we also have to impose $c_{L_x+1}^\dag=(-1)^{N+1}c_1^\dag$. Rewriting this in matrix form using Einstein's summation convention
\begin{equation}
 T c_i^\dag T^{-1}=M_{ij} c_j^\dag,
\end{equation}
we successively get
\begin{equation}
 T^\ell c_i^\dag T^{-\ell}=\left(M^\ell\right)_{ij} c_j^\dag
\end{equation}
and 
\begin{eqnarray}
\langle \sigma |T^\ell |\mu\rangle&=&\langle 0|c_{x_1} c_{x_2} \ldots c_{x_N} T^{\ell} c_{y_1}^\dag c_{y_2}^\dag \ldots c_{y_N}^\dag |0\rangle \\ 
&=&\langle 0|c_{x_1} c_{x_2} \ldots c_{x_N} (M^\ell)_{y_1 z_1}c_{z_1}^\dag (M^\ell)_{y_2 z_2}c_{z_2}^\dag \ldots (M^\ell)_{y_N z_N}c_{z_N}^\dag|0\rangle
\end{eqnarray}
By applying the Wick's theorem and using $\langle 0|c_x c_y^\dag|0\rangle=\delta_{xy}$, we finally get
\begin{equation}
 \langle \sigma |T^\ell |\mu\rangle=\det_{1\leq i,j\leq N} \left[(M^\ell)_{x_i y_j}\right]
\end{equation}
This type of result is referred to as the Karlin-McGregor\cite{KarlinMcGregor} or Lindstr\"om-Gessel-Viennot\cite{Lindstrom1973,GesselViennot1989} lemma in the mathematical literature. In the end, Eq.~(\ref{eq:tm}) reduces to a product of two determinants, which may be computed numerically.
\clearpage
\section[\;\;\;\;\;\;\;\;\;\;\;\;\;\;Dedekind-eta and Jacobi-Theta functions]{Dedekind-eta and Jacobi-Theta functions}
\label{sec:CFT_Jacobi}

For a given modulus $\tau$, we introduce the squared nome:
\begin{equation}
 q=e^{2i\pi \tau}
\end{equation}
The Dedekind eta function is defined as 
\begin{equation}\label{eq:eta_def}
 \eta(\tau)=q^{1/ 24}\prod_{k=1}^{\infty}\left(1-q^k\right),
\end{equation}
and the four Jacobi Theta functions are given by
\begin{eqnarray}\label{eq:theta1_def}
 \theta_1(z|\tau)&=&\sum_{n \in \mathbb{Z}}(-1)^{n-1/2}q^{\frac{1}{2}(n+1/2)^2}e^{(2n+1)iz}\\
 &=&2\eta(\tau)q^{1/6} \sin z\prod_{n=1}^{\infty} \left[1-2 \cos(2z) q^{n}+q^{2n}\right]
\end{eqnarray}
\begin{eqnarray}\label{eq:theta2_def}
 \theta_2(z|\tau)&=&\sum_{n \in \mathbb{Z}}q^{\frac{1}{2}(n+1/2)^2}e^{(2n+1)iz}\\
 &=&2\eta(\tau)q^{1/6} \cos z\prod_{n=1}^{\infty} \left[1+2 \cos(2z) q^{n}+q^{2n}\right]
\end{eqnarray}
\begin{eqnarray}\label{eq:theta3_def}
 \theta_3(z|\tau)&=&\sum_{n \in \mathbb{Z}}q^{\frac{1}{2}n^2}e^{2niz}\\
 &=&\eta(\tau)q^{-1/12}\prod_{n=1}^{\infty} \left[1+2 \cos(2z) q^{n-1/2}+q^{2n-1}\right]
\end{eqnarray}
\begin{eqnarray}\label{eq:theta4_def}
 \theta_4(z|\tau)&=&\sum_{n \in \mathbb{Z}}(-1)^{n}q^{\frac{1}{2}n^2}e^{2niz}\\
 &=&\eta(\tau)q^{-1/12}\prod_{n=1}^{\infty} \left[1-2 \cos(2z) q^{n-1/2}+q^{2n-1}\right]
\end{eqnarray}
To simplify the notations a bit, we also set
\begin{equation}\label{eq:thetas}
 \theta_\nu(\tau)=\theta_\nu (0|\tau)\quad,\quad \nu=1,2,3,4
\end{equation}
These functions obey the following nice ``modular'' transformation properties
\begin{eqnarray}
 \theta_2(\tau)&=&(-i \tau )^{-1/2}\,\theta_4(-1/\tau)\\\label{eq:modtheta3}
 \theta_3(\tau)&=&(-i \tau )^{-1/2}\,\theta_3(-1/\tau)\\\label{eq:modtheta4}
 \theta_4(\tau)&=&(-i \tau )^{-1/2}\,\theta_2(-1/\tau)\\\label{eq:modeta}
 \eta(\tau)&=&(-i \tau )^{-1/2}\,\eta(-1/\tau)
\end{eqnarray}

\section[\;\;\;\;\;\;\;\;\;\;\;\;\;\;CFT partition functions from the lattice dimer model]{Derivation of CFT partition function from the lattice dimer model}
\label{sec:dimers_exact}
 \subsection[\;\;\;\;\;\;\;\;\;\;\;\;\;\; Torus]{Torus}
The goal of this section is to recover the CFT partition function on the torus, using the exact solution for dimers in terms of transfer matrix, and keeping track of the winding numbers $W_x$ and $W_y$. To do that we need to solve a slightly more complicated problem than that in \ref{sec:lgv}, introducing a fugacity $b$ (resp. $b^{-1}$) for horizontal dimers whose left site belongs to the even (resp. odd) sublattice. We will also suppose $L_x$ and $L_y$ to be even. With this at hand the transfer matrix now satisfies
\begin{eqnarray}
T|0\rangle&=&|0\rangle\\
T c_{2j-1}^\dag T^{-1}&=&c_{2j}^\dag\\
T c_{2j}^\dag T^{-1}&=&b \,c_{2j}^\dag+c_{2j+1}^\dag +b^{-1}c_{2j+2}^\dag
\end{eqnarray}
The $b$'s provide information about $W_x$, while the number of fermions does that for $W_y$. 
Defining as before $T c_i^\dag T^{-1}=M_{ij} c_j^\dag$, diagonalizing $T$ amounts to diagonalizing $M$, which can be done in Fourier space, carefully taking into account the boundary condition on fermions
\begin{equation}
 c_{L_x+1}^\dag=(-1)^{\hat{N}+1}c_1^\dag\quad,\quad \hat{N}=\sum_{j=1}^{L_x} c^\dag_j c_j
\end{equation}
We write the number of dimer coverings in the $(W_x,W_y)$ sector as $Z_{W_x,W_y}$. Setting $b=e^{-\pi u_x/L_x}$, the winding generating function is given by
\begin{eqnarray}
 Z_{\rm torus}(u_x,u_y)&=&\sum_{W_x,W_y} Z_{W_x,W_y}e^{-\pi (W_x u_x+W_y u_y)}\\\label{eq:Z_tm}
 &=&{\rm Tr}\left[\displaystyle{e^{-\pi u_y(\hat{N}-L_x/2)/2}T^{L_y}}\right]
\end{eqnarray}
If we now denote by $d_k^\dag$ the set of fermionic operators that diagonalize the one-particle transfer matrix
\begin{equation}
 T d_k^\dag T^{-1}=\lambda_k d_k^\dag,
\end{equation}
the $\lambda_k$ are then the eigenvalues of $M$. Using this, the transfer matrix can be expressed as
\begin{equation}
 T=\prod_k \left(1+[\lambda_k -1]d_k^\dag d_k \right)
\end{equation}
Plugging this expression in Eq.~(\ref{eq:Z_tm}) we get\footnote{One can check easily that expanding simultaneously the first two terms in Eq.~(\ref{eq:dimers_torus}) generates all the eigenvalues of the transfer matrix in the even fermion sector. The same goes for the last two terms and the odd-fermions sector. }
\begin{equation}\label{eq:dimers_torus}
 Z_{\rm torus}(u_x,u_y)=\frac{1}{2}\left(Z_1^{+}+Z_1^{-}\right)+\frac{1}{2}\left(Z_0^{+}-Z_0^{-}\right)
\end{equation}
with
\begin{equation}\label{eq:torus01}
 Z_\nu^{\pm}=\prod_{k\in \Omega_{\nu}}\left[e^{\pi u_x /2}\pm e^{-\pi u_x/2}\lambda_k^{L_y}\right]
\end{equation}
and
\begin{eqnarray}\label{eq:lambda}
 \lambda_k&=&\cos k +\sqrt{1+\cos^2 k}\\
\Omega_{\nu}&=&\left\{\frac{(2m+\nu+i u_y) \pi}{L_x}\quad,\quad m=-L_x/2,\ldots,L_x/2-1\right\} 
\end{eqnarray}
It is also possible to rewrite $Z_\nu^\pm$ as 
\begin{equation}\label{eq:torus_cheb}
 Z_\nu^\pm=(-1)^{(\nu+1)L_y/2}\prod_{m=0}^{L_x/2-1}\left[e^{\pi u_x}+e^{-\pi u_x}\pm 2 T_{L_y}\left(i\cos \frac{(2m+\nu+iu_y)\pi}{L_x}\right)\right],
\end{equation}
where $T_m(x)=\cos(m\arccos x)$ is the $m-$th Chebyshev polynomial of the first kind. This expression allows for convenient numerical evaluations on large system sizes.

We now wish to extract the universal CFT partition function $\mathcal{Z}$, i.e the constant term in the asymptotic expansion of $Z$:
\begin{equation}
 Z_{\rm torus}(u_x,u_y)\sim A^{L_x L_y} B^{L_x} C^{L_y}\mathcal{Z}_{\rm torus}(u_x,u_y)
\end{equation}
A possible way to do so would be to combine Euler-Maclaurin expansions and majoration techniques, as has already been done for the honeycomb lattice\cite{Boutillier}. Here we obtain $\mathcal{Z}$ in a more heuristic (and non rigorous) manner, noticing that universal properties have to come from low energy excitation near the Fermi momenta $k_F=\pm \pi/2$. Linearizing $\ln(\lambda_k)$ around the two $k_F$ and extending the products in \ref{eq:dimers_torus} over all integers, allows to go get after a long calculation, very similar to that in \cite{Boutillier}:
\begin{equation}
 \mathcal{Z}_{\rm torus}(u_x,u_y)=\frac{\displaystyle{\theta_3(i\pi u_x/2|\tilde{\tau}/2)\,\theta_3(i\pi u_y/2|\tau/2)}}{\displaystyle{(2 \Im{\rm m}\, \tau)^{1/2}\,\eta(\tau)^2}}
\end{equation}
where $\tau=i L_y/L_x$ and $\tilde{\tau}=-1/\tau=iL_x/L_y$. 
\subsection[\;\;\;\;\;\;\;\;\;\;\;\;\;\; Cylinder]{Cylinder}
The cylinder geometry is slightly simpler than its torus counterpart, because only the windings along $y$ are allowed. To still express the partition function as a trace, we look at the transfer matrix $\tilde{T}$ acting on the configuration of rows of length $L_y$ with \emph{open} boundary conditions. This way, the number of fermions keeps track of the winding number $W_y$. The winding generating function can be expressed, after some algebra, as
\begin{eqnarray}
 Z_{\rm cyl}(u_y)&=&e^{-\pi (L_y/2)u_y}\prod_{m=1}^{L_y} \left(1+e^{\pi u_y}\mu_m^{L_x}\right)\\
 \mu_m&=&\cos \left(\frac{m \pi}{L_y+1}\right)+\sqrt{1+\cos^2\left(\frac{m \pi}{L_y+1}\right)}
\end{eqnarray}
Here the Fermi momentum is $k_F=\pi/2$, and the CFT partition function can be accessed using the linearization trick. We however have to distinguish between the even and odd $L_y$.
\paragraph{Even case:}
It is most convenient to rewrite $Z$ as
\begin{equation}
Z_{\rm cyl}(u_y)=\prod_{m=1}^{L_y/2} \lambda_m^{L_x}\times \left[\prod_{m=1}^{L_y/2} 1+2\cosh (\pi u_y)\mu_m^{-L_x}+\mu_m^{-2L_x}\right] 
\end{equation}
Combining the Euler-Maclaurin formula on the first product, and using the linearization procedure around $m=L_y/2$, we get
\begin{equation}
 \mathcal{Z}_{\rm cyl}(u_y)= e^{\frac{\pi L_x}{24 L_y}}\prod_{p=1}^{\infty}\left[1+2 \cosh (\pi u_y)e^{-\frac{\pi L_x}{L_y}(p-1/2)}+e^{-2\frac{\pi L_x}{L_y}(p-1/2)}\right]
\end{equation}
which gives, using the infinite product representation of the Jacobi Theta function,
\begin{equation}
 \mathcal{Z}_{\rm cyl}(u_y)=\frac{\theta_3(i \pi u_y/2|\tilde{\tau}/2)}{\eta(\tilde{\tau}/2)}
\end{equation}
We recover the Eq.~(\ref{eq:cylinder_dd}), upon setting $u_y=0$ and performing a modular transformation. 
\paragraph{Odd case:}
We use the same method, but care must be taken because of the presence of a zero-mode for $m=(L_y+1)/2$ We have
\begin{equation}
Z_{\rm cyl}(u_y)=2\cosh(\pi u_y/2)\prod_{m=1}^{(L_y-1)/2} \lambda_m^{L_x}\times \left[\prod_{m=1}^{(L_y-1)/2} 1+2\cosh (\pi u_y)\mu_m^{L_x}+\mu_m^{2L_x}\right] 
\end{equation}
which yields after linearization
\begin{equation}
 \mathcal{Z}_{\rm cyl}(u_y)=2 e^{-\frac{\pi L_x}{12 L_y}}\cosh(\pi u_y/2)\prod_{p=1}^{\infty}\left[1+2 \cosh (\pi u_y)e^{-\frac{\pi L_x}{L_y}p}+e^{-2\frac{\pi L_x}{L_y}p}\right]
\end{equation}
In the end we obtain:
\begin{equation}
 \mathcal{Z}_{\rm cyl}(u_y)=\frac{\theta_2(i \pi u_y/2|\tilde{\tau}/2)}{\eta(\tilde{\tau}/2)}
\end{equation}
and once again recover the CFT result (\ref{eq:cylinder_ddprime}) after setting $u_y=0$ and modular transformation. 
\subsection[\;\;\;\;\;\;\;\;\;\;\;\;\;\; Zero-winding sectors]{Zero-winding sectors}
With the winding generating functions at hand, the calculation of $s_n(y,\tau)$ in the $W=(0,0)$ winding sector becomes straightforward. For example we have in the odd case (recall $\tau=iL_y/L_x$ and $\tilde{\tau}=-1/\tau$)
\begin{equation}
 \underbrace{\mathcal{Z}_{\rm cyl}^{DD^\prime}(y\tau)\mathcal{Z}_{\rm cyl}^{DD^\prime}((1-y)\tau)}_{W=(0,0)}=\frac{\theta_2\Big(0\left|\frac{\tilde{\tau}}{2y(1-y)}\right)}{\eta(\frac{\tilde{\tau}}{2y})\eta(\frac{\tilde{\tau}}{2(1-y)})},
\end{equation}
where we have selected the constant term in the product of the two theta functions.  
This allows to recover Eq.~(\ref{eq:cft_w0_gen}) and Eq.~(\ref{eq:cft_w0_gen2}) after once again performing a modular transformation. 
 \section*{References}
\providecommand{\href}[2]{#2}\begingroup\raggedright\endgroup

\end{document}